\documentclass[12pt]{article}

\usepackage{times}
\usepackage{authblk}
\usepackage{graphicx}
\graphicspath{ {images} }
\usepackage[utf8]{inputenc}
\usepackage[T1]{fontenc}
\usepackage{amsmath}
\usepackage{amstext}
\usepackage{amsopn}
\usepackage{ulem}
\usepackage{ amsbsy}
\usepackage{amsfonts}
\usepackage{amssymb}
\usepackage{graphicx}
\usepackage[nottoc]{tocbibind}
\usepackage{braket}
\usepackage{physics}
\usepackage{booktabs}
\usepackage{subcaption}
\usepackage{color}
\definecolor{prdblue}{rgb}{0.133,0.118,0.698}
\usepackage[colorlinks=true, pdfstartview=FitV, linkcolor=prdblue, 
citecolor= prdblue, urlcolor=prdblue]{hyperref}

\title{Coherence of oscillations in matter and 
supernova neutrinos}

\author{
\\
\normalsize{}
\\
\normalsize{}
}

\author{{Yago} P. Porto-Silva$^{\ast}$, Alexei Yu. Smirnov$^\dagger$
\\
\normalsize{ $^\ast$ Instituto de F{\'i}sica Gleb Wataghin - UNICAMP, 13083-859,
Campinas, S\~ao Paulo, Brazil}
\\
\normalsize{$^\ast$ $^\dagger$ Max-Planck-Institut 
f\"ur Kernphysik, 69117 Heidelberg, Germany}
\\

\normalsize{E-mail: $^\ast$porto@mpi-hd.mpg.de, 
$^\dagger$smirnov@mpi-hd.mpg.de }
}

\date{}

\begin{document}
\maketitle

\begin{abstract}


We study the propagation coherence for neutrino oscillations in
media with different density profiles.
For each profile, we find the dependence of the
coherence length, $L_{coh}$, on neutrino energy and
address the issue of correspondence
of results in the distance and energy-momentum representations.
The key new feature in matter is existence of
energy ranges with enhanced coherence around the energies
$E_0$ of "infinite coherence" at which $L_{coh} \rightarrow \infty$.
In the configuration space, the infinite coherence corresponds
to equality of the (effective) group velocities of the eigenstates.
In constant density medium, there is a unique $E_0$, which coincides with the MSW resonance energy of oscillations of mass states and is close to the MSW resonance energy of flavor states.
In the case of massless neutrinos or negligible masses in a very dense
medium
the coherence persists continuously.
In the adiabatic case, the infinite coherence is realized for periodic
density change. Adiabaticity violation changes the shape factors of
the wave packets (WPs) and leads to their spread. In a medium with sharp density
changes (jumps), splitting of the eigenstates occurs at crossing
of each jump. We study the increase of the coherence length
in a single jump and periodic density jumps - castle-wall (CW) profiles.
For the CW profile, there are several $E_0$ corresponding to parametric resonances. We outlined applications of the results for supernova neutrinos.
In particular, we show that coherence between two shock wave fronts leads to observable oscillation effects, and our analysis suggests that the decoherence can be irrelevant for flavor transformations in the central parts of collapsing stars.
\end{abstract}

\section{Introduction}

Propagation coherence is the condition for realization of
oscillations - the interference effects that lead to
time - distance periodic variations of observables \cite{Kayser:1981ye,Kiers:1995zj,Giunti:1997wq,Grimus:1998uh,Giunti:2002xg,Akhmedov:2009rb,Akhmedov:2010ms,Akhmedov:2012uu,Akhmedov:2019iyt}. 
Correspondingly, decoherence means the disappearance of the interference 
and oscillatory pattern. 

In the configuration space, the decoherence is produced by 
relative shift, and eventually, separation of the wave 
packets (WPs) that correspond to the eigenstates of propagation. The coherence length was defined as the distance at which
the wave packets are shifted with respect to 
each other (separated) by the size of the wave packet $\sigma_x$, or equivalently, when the interference term in the probability is suppressed by a factor $1/e$.
In the energy-momentum space, the decoherence is due to 
averaging of the oscillation phase over the energy interval of the energy uncertainty in a setup. 
It was checked that in vacuum and uniform matter, the two considerations 
are equivalent and lead to the same value of coherence length \cite{Mikheev:1986wj,Mikheev:1987jp,Kersten:2015kio}.

The loss of the propagation coherence does not lead to the
loss of information and can be restored under certain conditions. In the configuration space, the restoration requires a long enough coherence time of detection. In the energy-momentum 
space it requires very good energy resolution. Irreversible loss of coherence occurs if neutrinos are treated as open quantum systems \cite{Benatti:2000ph}.

Mostly the coherence was studied for vacuum oscillation \cite{Kayser:1981ye,Kiers:1995zj,Giunti:1997wq,Grimus:1998uh,Giunti:2002xg,Akhmedov:2009rb,Akhmedov:2010ms,Akhmedov:2012uu,Akhmedov:2019iyt}. 
In matter with constant (or adiabatically changing) density
it was considered in \cite{Mikheev:1987jp,Mikheyev:1989dy,PhysRevD.41.2379,Peltoniemi:2000nw,Kersten:2015kio}. It was noticed that 
at certain energy (close to the MSW resonance energy), the difference of group velocities of the WPs vanishes in media with constant density and the 
coherence length becomes infinite \cite{Mikheev:1986wj,Mikheev:1987jp,Kersten:2015kio}. If density varies with distance monotonously, then at a specific point, $x_0$, near the MSW resonance, the difference of group velocities changes sign, and the shift that happened at $x<x_0$ can be undone at $x>x_0$, so that the packets overlap can be restored \cite{Mikheev:1987jp,Mikheyev:1989dy,Kersten:2015kio}. Decoherence affects resonant oscillations of solar and supernova neutrinos \cite{PhysRevD.41.2379} and, in the case of supernova, decoherence happens before the resonance region. However, the effects of collective oscillations were not taken into account in \cite{PhysRevD.41.2379}. In ref.~\cite{Peltoniemi:2000nw}, a formal solution for the neutrino equation of motion in general profile including non-adiabaticity was found. Based on it, the equivalence between $x$- and $E$-spaces was shown, at least for situations with $\sigma_E \ll E$. Limitations of the WP treatment were also discussed; in particular, as $\sigma_E$ increases, negative energy components of the WPs become relevant. This can be important in situations with $\sigma_E \sim E$.

The coherence in the inner parts of supernova (SN)
is of special interest \cite{Kersten:2013fba}. The reason is that the wave packets of SN neutrinos are very short, $\sigma_E \sim E$, suggesting that decoherence can occur at smaller scales than
the scales of possible collective oscillation phenomena \cite{Kersten:2015kio,Akhmedov:2017mcc}. 
Furthermore, effectively, the problem is non-linear \cite{Hansen:2018apu}. 
Therefore the question arises whether the loss of coherence destroys or not the collective phenomena \cite{Hansen:2019iop}. 

In this paper, we consider coherence and decoherence in matter in detail. We revisit the case of the infinite coherence length in matter of constant density and give its physics interpretation. We show that in matter dominated regime, decoherence is determined by vacuum parameters. We comment on the fact that oscillations of massless neutrinos do not decohere. In the adiabatic case $L_{coh} \rightarrow \infty$ is possible if a density profile has periodic modulations.
We then consider decoherence in the presence of adiabaticity violation. That includes decoherence in matter with single density 
jump which can be relevant for propagation through the shock wavefront in SN \cite{Kersten:2015kio}, and in the castle-wall profile \cite{Akhmedov:1988kd,Krastev:1989ix,Akhmedov:1998ui,Akhmedov:1999ty,Akhmedov:1999va}. The latter can give some idea about the decoherence of collective oscillations in the inner parts of supernova \cite{Hansen:2018apu}. The castle-wall profile is the simplest example because it admits analytical treatment and allows us to draw some conclusions about the relation between parametric resonances and coherence length. 

Decoherence in the configuration ($x$-space) is equivalent to averaging of oscillation phase in energy or $E$-space for stationary situations \cite{Stodolsky:1998tc}. Therefore, for each matter profile, we elaborate on the
issue of equivalence of results in the $x$- and $E$-spaces. In particular, for the castle-wall profile, we show that the modification of shape factors in $x$-space due to the density jumps do not alter its Fourier transform and the initial energy spectrum of the WPs is preserved.

The paper is organized as follows: in section~\ref{coherence-lengths} we study the coherence length in vacuum and matter with constant and adiabatically varying density. We emphasize the existence of energies with infinite coherence length and give interpretation of $L_{coh} \rightarrow \infty$ in position and energy spaces.
In section~\ref{coherence and adiabaticity violation}, we analyze coherence in the case of maximal adiabaticity violation.
Section~\ref{parametric-oscillations} presents the study of coherence in the castle-wall profile.
Section~\ref{applications} is devoted to the applications of our results to supernova neutrinos.
In section~\ref{conclusions} we present our conclusions.

\section{Coherence in matter with constant and adiabatically changing density} 
\label{coherence-lengths}

In what follows for simplicity, we will consider the coherence in two neutrino system; 
the generalization to the case of three neutrinos is straightforward. 

\subsection{Coherence of oscillations in matter}

In the energy-momentum space, the decoherence is related 
to averaging of the oscillation phase $\phi$ or 
the oscillatory (interference) term in probability $P_{int}$ 
over the energy uncertainty, $\sigma_E$, of a set up\footnote{There is subtle aspect related to averaging 
of $\phi$ and $P_{int}$ which we will discuss later.}.
$\sigma_E$ gives the width of the wave packet (WP), and 
in general, it is determined by the energy uncertainty at the production 
and detection, thus including a possibility 
of restoration of the coherence at the detection: 
$1/\sigma_E = 1/\sigma_E^{prod} + 1/\sigma_E^{det}$ \cite{Kayser:1981ye,Kiers:1995zj,Giunti:1997wq,Grimus:1998uh,Giunti:2002xg,Akhmedov:2009rb,Akhmedov:2010ms,Peltoniemi:2000nw,Hollenberg:2011tc,Kersten:2015kio}.

In the case of uniform medium (vacuum, constant density matter) the 
eigenstates and eigenvalues of the Hamiltonian of propagation 
$H_{im}$ ($i = 1,\, 2$) are well defined \cite{Wolfenstein:1977ue,Mikheev:1986wj,Mikheev:1987jp,Mikheyev:1989dy}. The difference of the eigenvalues, $\Delta H_m \equiv H_{2m} - H_{1m}$, determines the oscillation phase
acquired along the distance $L$: 
\begin{equation} 
\label{mat-phase} \nonumber
\phi_m = \Delta H_m L.
\end{equation}
For a fixed $L$, variation of the phase with change of the neutrino 
energy $E$ in the interval $E \pm \sigma_E$ equals 
\begin{equation} 
\label{expansion}
\Delta \phi_m = \left( 2 \sigma_E 
\frac{d \Delta H_m}{d E}
+ \frac{\sigma_E^3}{3}\frac{d^3 \Delta H_m}{d E^3} 
+... \right) L. 
\end{equation}
It increases linearly with $L$ and we define the coherence 
length, $L_{coh}$, 
as the length at which the variation of oscillation phase
becomes $2\pi$: 
\begin{equation}
\label{cohlength}
|\Delta \phi_m (L_{coh}^m)|= 2 \pi.
\end{equation}
This condition means that averaging of the oscillatory dependence of the probability, which is a measure of interference, 
becomes substantial. 
Using the first term of the expansion (\ref{expansion}) 
we obtain from the condition (\ref{cohlength}) 
\begin{equation} 
\label{const-density}
L^m_{coh} = \frac{\pi}{\sigma_E} 
\left| \frac{d \Delta H_m}{d E} \right|^{-1}. 
\end{equation}
At 
\begin{equation} 
\label{infcoh}
\frac{d \Delta H_m (E)}{d E} = 0 
\end{equation}
the coherence length becomes infinite \cite{Mikheev:1987jp,Mikheyev:1989dy,Kersten:2015kio}. The next terms in the expansions (\ref{expansion}) shift the pole in (\ref{const-density})
but do not eliminate it. The shift due to higher order terms is suppressed if
$\sigma_E/E \ll 1$.

In vacuum 
\begin{equation} \nonumber
\Delta H = \frac{\Delta m^2}{2 E},
\end{equation}
where $\Delta m^2 \equiv m_2^2 -m_1^2$ is the mass square difference.
Consequently, Eq.~(\ref{const-density}) gives
\begin{equation} 
\label{L-coh}
L_{coh} = 
\frac{\pi}{\sigma_E}\frac{2 E^2}{\Delta m^2} = 
l_{\nu} \frac{E}{2 \sigma_E}, 
\end{equation}
with $l_{\nu} = 4 \pi E / \Delta m^2$ being 
the vacuum oscillation length. 
Notice that $L_{coh}\rightarrow \infty$ in the limits 
$E \rightarrow \infty$ or $\Delta m^2 \rightarrow 0$. 

In matter with constant 
density \cite{Kersten:2015kio}, the difference 
of eigenvalues for a given matter potential is given by 
\begin{equation} 
\label{split-eigenvalues}
\Delta H_m = \frac{\Delta m^2}{2E} 
\sqrt{\left(c_{2\theta} - \frac{2EV}{\Delta m^2} \right)^2 
+ s^2_{2\theta}} \, ,
\end{equation}
where $\theta$ is the vacuum mixing angle and $c_{2 \theta} \equiv \cos 2 \theta$ and $s_{2 \theta} \equiv \sin 2\theta$. The derivative of (\ref{split-eigenvalues}) equals 
\begin{equation}
\label{eq:partder}
\frac{d \Delta H_m}{d E} = 
\frac{\Delta m^2}{2E^2} 
\frac{ \left(\frac{\Delta m^2}{2 E} - 
V \cos 2 \theta \right)}{\Delta H_m}.
\end{equation}
Consequently, according to Eq.~(\ref{const-density}) 
the coherence length in matter equals 
\begin{equation} \nonumber
\label{const-density1}
L^m_{coh}= 
\frac{L_{coh}}{l_m} \frac{2 \pi}{\left|\frac{\Delta m^2}{2 E} 
- V \cos 2 \theta \right|} = L_{coh} 
\left(1 -c_{2\theta} \frac{2VE}{\Delta m^2} \right)^{-1} 
\sqrt{\left( c_{2\theta} - \frac{2VE}{\Delta m^2} \right)^2 + s^2_{2 \theta}} ,
\end{equation}
with $L_{coh}$ given in (\ref{L-coh}) and $l_m=2 \pi / \Delta H_m$ 
being the oscillation length in matter. 

Let us consider dependence of $L^m_{coh}$ on energy. 
Its salient feature is existence of the pole at 
\begin{equation} \nonumber
\label{eq-E0}
 E_0=\frac{\Delta m^2}{2 V \cos 2 \theta}. 
\end{equation}
That is, the infinite coherence length 
$L^m_{coh} \rightarrow \infty$ is realized 
at finite energy $E_0$, in contrast 
to the vacuum case. $E_0$ is related to the MSW resonance energy, $E_R =\Delta m^2 \cos 2\theta/2 V $, as 
\begin{equation} 
\label{MSW}
E_0 = \frac{E_R}{\cos^2 2\theta}. 
\end{equation}
Thus, $E_0 > E_R$ and it can be very close to $E_R$ in case of small mixing. Since the width of the MSW resonance peak is $2\tan 2 \theta E_R \approx 2\sin 2\theta E_R$, the pole is within the peak. 
Therefore, at $E_0$, the transition probability is large. 

\begin{figure} 
\centering
\includegraphics[width=0.8\linewidth]{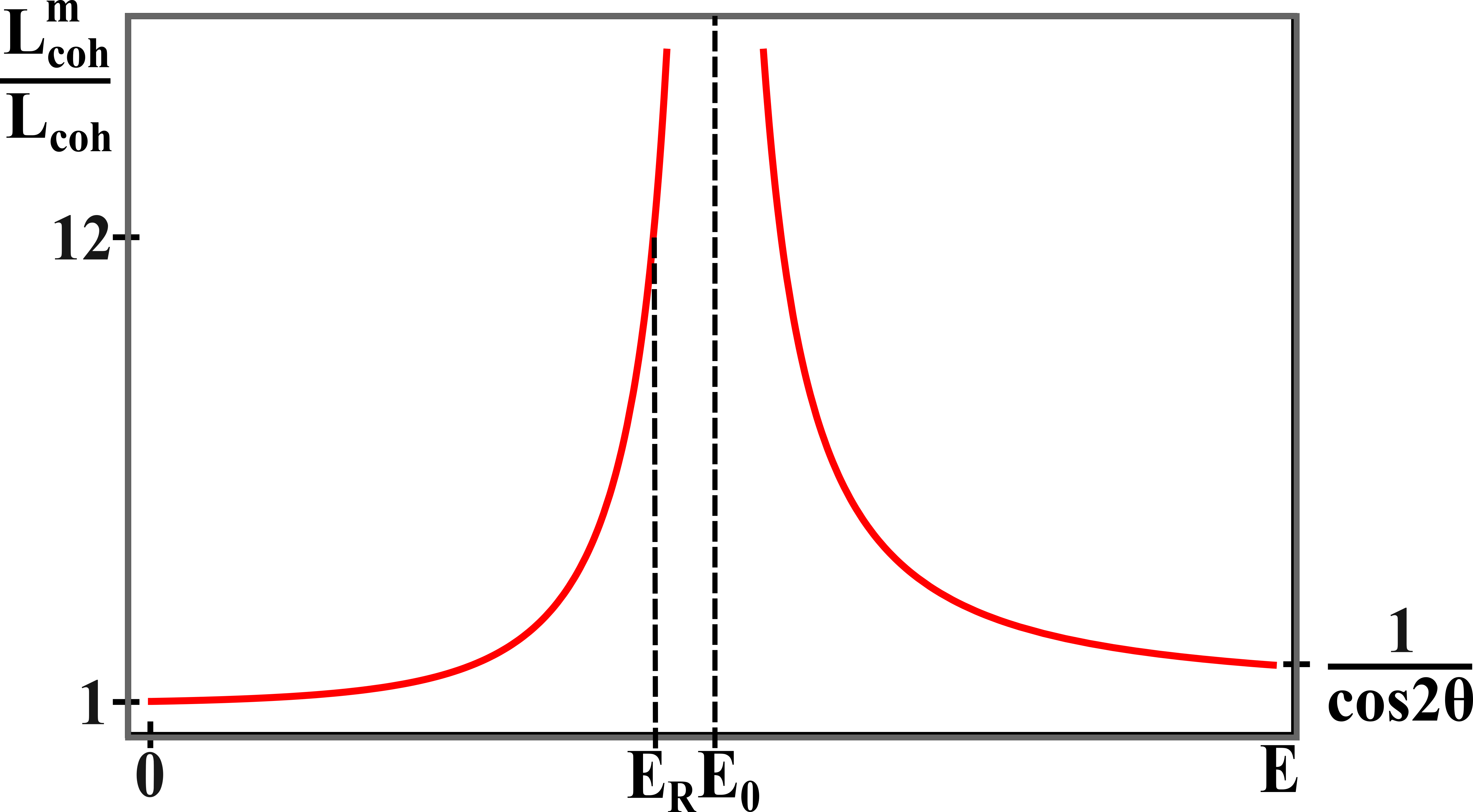}
\caption{Dependence of the ratio $L^m_{coh}/L_{coh}$ (\ref{ratio}) 
on energy for $\theta = \theta_{13} = 8.5^\circ $. The ratio diverges at $E_0$, and $E_R$ is the MSW resonance energy. 
}
\label{fig-E0}
\end{figure}

In terms of $E_0$ or $x \equiv E/E_0$ the derivative (\ref{eq:partder}) 
can be rewritten as 
\begin{equation} \nonumber
\label{eq:intermse0}
\frac{d \Delta H_m}{d E} = 
\frac{2\pi}{E l_\nu} 
\frac{1 - x}{\sqrt{(1 - x)^2 + x^2\tan^2{2\theta} }}. 
\end{equation}
Consequently, the ratio of the coherence lengths 
in matter and in vacuum (\ref{const-density}) becomes 
\begin{equation} 
\label{ratio}
 \frac{L^m_{coh}}{L_{coh}}=\sqrt{1+ 
\frac{\tan^2 2 \theta}{\left(1-\frac{E_0}{E} \right)^2}}.
\end{equation}
The ratio Eq.~(\ref{ratio}) as function of energy 
is shown in fig.~\ref{fig-E0}. It
has the universal form which depends on the vacuum mixing only. 
The value $L^m_{coh}/L_{coh} = 2$ corresponds to energies at both
sides of the peak $E=E_0 \left(1 \pm \tan 2 \theta/\sqrt{3}\right)^{-1}$.
Thus, for small $\theta$ the quantity $\Delta E = (2/\sqrt{3})\tan 2\theta E_0$
characterizes the width of the peak.

At the MSW resonance, $E=E_R$, the ratio (\ref{ratio}) equals
\begin{equation} 
\label{ratio-resonance}
\frac{L^m_{coh}}{L_{coh}}=\frac{1}{\sin 2 \theta}.
\end{equation}
That is, the coherence length Eq.~(\ref{ratio-resonance})
is enhanced in comparison to the vacuum length. 
For example, if $\theta=\theta_{13}=8.5^\circ$, 
we have $L^m_{coh} \approx 12 L_{coh}$. 
In the resonance the oscillation length increases 
by the same amount: $l_m^R = l_\nu/ \sin 2\theta$.

For $E \ll E_0$, {\it i.e.} in the vacuum dominance case,
we have $L^m_{coh} \approx L_{coh}$. 
Interestingly, for $E \gg E_0$, (matter dominated region)
\begin{equation} 
L^m_{coh} \approx \frac{L_{coh}}{\cos 2 \theta}. 
\label{matter-domination}
\end{equation}
That is, $L_{coh}^m$ is determined by the vacuum parameters, and for small $\theta$ it approaches the vacuum $L_{coh}$ again. 
This has important implications for supernova neutrinos. 

The depth of oscillations in matter given by
\begin{equation} 
\label{sine-mixing}
\sin^2 2 \theta_m = \frac{\sin^2 2 \theta}{\left( \cos 2\theta 
- \frac{2EV}{\Delta m^2} \right)^2 + \sin^2 2\theta}
\end{equation}
becomes at $E=E_0$
\begin{equation} \nonumber
 \sin^2 2 \theta_m =\cos^2 \theta.
\end{equation}
For $\theta=8.5^0$ the depth is close to maximum: 
$\sin^2 2 \theta_m \approx 0.98$. 
Therefore, in matter, strong flavor transitions can be preserved 
from decoherence over large distances $L$ from the source to the detector.





\subsection{Coherence in the configuration space}

In the configuration space, the loss of propagation coherence 
is associated with relative shift and eventually separation 
of the wave packets of the eigenstates. 
The difference of group velocities 
of the eigenstates is given by 
\begin{equation} 
\label{Delta-v}
 \Delta v_m=-\frac{d \Delta H_m}{d E}.
\end{equation}
Then spatial separation (relative shift) 
$x_{shift}$ of the packets after propagating a distance $L$ equals
\begin{equation} \nonumber
\label{velocity}
x_{shift}(L) =\int_{0}^{L} \Delta v_m dx = 
-\int_{0}^{L} \frac{d \Delta H_m}{d E} dx.
\end{equation}
The coherence length $L^m_{coh}$ can be defined as the distance at which 
the separation equals the spatial size of the packets $\sigma_x$:
$|x_{shift}(L^m_{coh})| \approx \sigma_x$. At this point, the overlap of the packets becomes small. 
In the case of constant density this condition gives 
\begin{equation} 
\label{velocity1}
L^m_{coh} = \sigma_x 
\left| \frac{d \Delta H_m}{d E} \right|^{-1}.
\end{equation}
Since $\sigma_x = 2\pi/(2 \sigma_E)$ 
(recall that $\sigma_E \approx \sigma_p$ is the half-width), 
the expression (\ref{velocity1}) coincides with expression 
for the length in the energy representation (\ref{const-density}). 
This shows the equivalence of the results in the 
configuration and the energy-momentum spaces. 

According to Eq.~(\ref{Delta-v}) at $E=E_0$ 
the eigenstates in matter have equal group 
velocities and therefore do not separate. As a result, 
coherence is maintained during infinite time: 
$L^m_{coh} \rightarrow \infty$. 
Thus, in the configuration space the condition for infinite coherence 
is 
\begin{equation} 
\label{infcoh-conf}
v_1 = v_2. 
\end{equation} 

The infinite coherence energy $E_0$ coincides
with the resonance energy of oscillations
of the mass eigenstates in matter
$\nu_1 \leftrightarrow \nu_2$ \cite{Mikheyev:1989dy}. This is not accidental.
The origin of the energy dependence of phase,
the difference of group velocities
and therefore decoherence is the mass states.
The resonance means that at $E_0$ the mass states $\nu_1$ and $\nu_2$
are maximally mixed in the eigenstates $\nu_i^m$.
That is, $\nu_1^m$ and $\nu_2^m$
both contain equal admixtures of $\nu_i$. Consequently,
the group velocities of $\nu_1^m$ and $\nu_2^m$ should be equal.
In other words, in a given flavor state
the components $\nu_1$ and $\nu_2$ oscillate
with maximal depth $\nu_1 \rightarrow \nu_2
\rightarrow \nu_1 \rightarrow \nu_2$,
which compensate separation.

The derivative of the oscillation length in matter equals
$$
\frac{d l_m}{dE} = - \frac{2\pi}{(\Delta H_m)^2}
\frac{d\Delta H_m}{dE}.
$$
Therefore the condition of infinite coherence (\ref{infcoh}) coincides
with the extremum of oscillation length.
For a given potential $V$,
$l_m^{max} = 2\pi/(V \sin 2\theta)$, which is larger than
the length in the MSW resonance:
$l_m^{R} = 2\pi/(V \tan 2\theta)$, and the difference
is substantial for large vacuum mixing.

Notice that decoherence is related to
uncertainty in energy on a certain interval of energy.
Therefore infinite coherence energy should refer to the
average energy in a wave packet.

Notice that the equivalence of results in two representations 
originates from the fact that 
the same quantity $d \Delta H_m /d E$ 
determines the change of the oscillation phase with energy 
(in $E-p$ space) on the one hand side, and the difference of the 
group velocities (in $x$-space) on the other hand.

\subsection{Correlation of the phase difference and delay}

The separation between the WP of eigenstates, $t \approx \Delta v X$,
is proportional to their phase difference (oscillation phase), 
$\phi = \Delta H_m X$: 
\begin{equation}
\label{eq:phase-delay}
t = g(E, V) \phi, 
\end{equation}
where according to (\ref{eq:partder})
\begin{equation} \nonumber
\label{eq:gfunc}
g(E, V) = \frac{\Delta v}{\Delta H_m} = 
\frac{1}{\Delta H_m}{2E^2} = 
\frac{1}{E} 
\frac{1 - \frac{2 V E c_{2\theta}}{\Delta m^2}} 
{\left(c_{2\theta} - \frac{2VE}{\Delta m^2}\right)^2 + s^2_{2\theta}}.
\end{equation}
In terms of infinite coherence energy, $E_0$, it can be rewritten as
\begin{equation} \nonumber
\label{eq:gfunc2}
g(E, V) = \frac{2 V c_{2\theta}}{\Delta m^2} 
\frac{\frac{E_0}{E} - 1} 
{\left(c_{2\theta} - \frac{E}{E_0 c_{2\theta}} \right)^2 
+ s^2_{2\theta}}.
\end{equation}
For $E \ll E_0$ (vacuum case)
$$
g(E, V) = \frac{1}{E}.
$$
At $E = E_0$ (near the MSW resonance): 
$g(E, V) = 0$ - there is no loss of coherence, 
and the oscillation phase, being non-zero, is suppressed by mixing: 
$$
\phi = \frac{\Delta m^2 X}{2E_0} \tan 2\theta = 
\phi_{vac}(E_0) \tan 2\theta. 
$$
For $E \gg E_0$ (matter dominated case):
$$
g(E, V) \approx - \frac{c_{2\theta} \Delta m^2}{2V E^2}= - \frac{ E_0 c_{2 \theta}^2}{ E^2}.
$$
Here also the relative separation is strongly 
suppressed, while oscillations proceed with large phase 
$\phi \approx V X$. 

Notice also that the relative separation changes the sign at $E_0$. Therefore in layers $a$ and $b$ with 
two different densities one may have opposite sign 
of a delay and also 
$t^a = -t^b$ (see below). The phases do not change the sign and are
always positive.

\subsection{Physics of $L^m_{coh} \rightarrow \infty$} 
\label{interpretation}

Here we present an interpretation of the divergence of 
$L^m_{coh}$ at certain energy $E_0$. 
The transition probability is given by
\begin{equation} \nonumber
\label{generic-prob}
P_{\alpha \beta}(\theta_m,\phi_m ) =
\frac{1}{2}\sin^2 2 \theta_m \left(1 - \cos \phi_m \right).
\end{equation}
It should be averaged over the energy interval $2\sigma_E$. 
The change of $\sin^2 2 \theta_m$ with energy in the interval 
$E_0 \pm \sigma_E$ equals
\begin{equation}\nonumber
 \left| 2 \sigma_E \frac{d}{d E}\sin^2 2 \theta_m \right|=2 \sigma_E \frac{V c_{2 \theta}}{\Delta m^2}=\frac{\sigma_E}{E_0} \ll 1,
\end{equation}
where we have taken into account that at $E_0$ the mixing parameters are $\sin^2 2 \theta_m=\cos^2 \theta$ and $\cos^2 2 \theta_m=\sin^2 \theta$. Therefore 
$\sin^2 2 \theta_m$ can be put out of the averaging integral at $E \sim E_0$. 
Furthermore, to simplify consideration, instead of averaging over $E$ we will 
average $P_{\alpha \beta}$ immediately over the phase $\phi_m$ 
in the interval $2\delta \phi_m$ determined by 
\begin{equation}
2 \delta \phi_m = \frac{d \phi_m }{d E} 2\sigma_E
= \frac{d \Delta H_m}{d E} L 2\sigma_E. 
\label{pm-int}
\end{equation}
The averaging of the oscillatory term yields
\begin{equation} \nonumber
\label{average}
\frac{1}{2 \delta \phi_m} \int_{\phi_m - \delta \phi_m}^{\phi_m
+ \delta \phi_m} 
d\phi_m' P_{\alpha \beta}(\phi_m') = \frac{1}{2}
\sin^2 2 \theta_m (E_0) 
\left[1 - D(\delta \phi_m) \cos \phi_m \right], 
\end{equation}
where 
\begin{equation} \nonumber
\label{decoh}
D(\delta \phi_m) \equiv \frac{\sin \delta \phi_m}{\delta \phi_m}
\end{equation}
is the decoherence factor which describes suppression of 
the interference term.

According to (\ref{pm-int}) for $E = E_0$ the variation of phase 
is $\delta \phi_m = 0$, 
and consequently, $D(\delta \phi_m) \rightarrow 1$. 
The averaging effect is small in the region around $E_0$. 
Indeed, $D(\delta \phi_m)$ has a peak centered at $\delta \phi_m = 0$. 
The half-width at nearly half of maximum, 
$D(\delta \phi_m) = 0.5$ corresponds to 
\begin{equation} \label{rounding}
|\delta \phi_m| \approx 1.89. 
\end{equation}
Consequently, the half-width of the peak in the energy scale, $\Gamma$, 
is determined by the condition 
\begin{equation} \label{Gamma}\nonumber
|\delta \phi_m (E_0 + \Gamma)| = 2
\end{equation}
(rounding (\ref{rounding}) to $2$).
Using this value 
and expressions in (\ref{pm-int}) and (\ref{eq:partder}), 
we find the relative half-width of the peak:
\begin{equation} 
\label{exact-peak}
 \frac{\Gamma}{E_0} = 
\frac{ \tan 2 \theta }{\sqrt{\pi^2 d^2 - 1}},
\end{equation}
where 
\begin{equation}\nonumber
d \equiv \frac{\sigma_E}{E_0} \frac{L}{l_{\nu}}. 
\end{equation}
Notice that (\ref{exact-peak}) requires $d \pi > 1$, 
on explicitly 
$L > l_{\nu}E/ (\pi \sigma_E) \approx L_{coh}$.
For $d \pi \rightarrow 1$, the width $\Gamma \rightarrow \infty$, 
which means that coherence is well satisfied for all the energies. 
Moreover, the peak disappears (half a height does not exist). 

For $\pi d \gg 1$ the Eq.~(\ref{exact-peak}) becomes
\begin{equation} 
\label{exact-peak1}
\frac{\Gamma}{E_0} = \frac{ \tan 2 \theta }{\pi d} = 
\tan 2 \theta \frac{E_0}{\sigma_E} \frac{l_{\nu}}{\pi L}. 
\end{equation}
The larger $L$, as well as $\sigma_E$, the narrower the peak.
Also, according to (\ref{exact-peak1}) the width of the peak increases with energy:
\begin{equation} \nonumber
\label{propto-peak}
 \frac{\Gamma}{E_0} \propto E_0^2. 
\end{equation}



\subsection{Adiabatic evolution and infinite coherence}
In the case of slow adiabatic density change \cite{Mikheev:1986wj,Mikheev:1987jp,Mikheyev:1989dy}, one can
introduce the instantaneous eigenstates and eigenvalues
$H_{im}(x)$ and their difference $\Delta H_m(x)$,
which are well-defined quantities.
The adiabatic oscillation phase is given by the integral
\begin{equation}
\label{adiab-phase}
\phi_m^{ad}(L) = \int_0^{L} dx \Delta H_m(N_e(x)).
\end{equation}
A change of the phase with the energy in the interval
$2\sigma_E$ equals
\begin{equation}
\label{adiab-diff}\nonumber
\Delta \phi_m^{ad}(L) = 2 \sigma_E \int_0^{L} dx
\frac{d}{d E}\Delta H_m(x),
\end{equation}
where we permuted the intergration over $x$ and differentiation
over $E$.
Then the condition for decoherence is
$\Delta \phi_m (L^m_{coh}) = 2\pi$, or explicitly,
\begin{equation}
\label{cond-var-density}\nonumber
\left| \int_0^{L^m_{coh}} dx \frac{d}{d E}
\Delta H_m(x)\right| = \frac{\pi}{\sigma_E},
\end{equation}
and $d \Delta H_m(x)/d E$ is given in (\ref{eq:partder}).

In the configuration space the adiabatic evolution means
that there are no transitions between the eigenstates.
Therefore a propagation is described by two wave packets,
which do not change shape factors as in the constant density case.
Separation of the WP equals
\begin{equation}
\label{adiab-sep}
\Delta x (L) = \int_0^{L} dx
\frac{d}{d E} \Delta H_m(x),
\end{equation}
and from (\ref{adiab-phase}), (\ref{adiab-sep}) we obtain
\begin{equation}\nonumber
\label{adiab-equal}
\frac{d\phi_m^{ad}}{d E} = \Delta x (L),
\end{equation}
which is the basis of equivalence of results in the $x$- and $E$-representations. In particular,
the infinite coherence condition, $d\phi_m^{ad} /dE = 0$, correspons to
zero separation.

Let us consider coherence for different density (potential) profiles.
For monotonous change of the potential the infinite
$L_{coh}^m$ can not be realized. Indeed, for a given $E$ the pole
condition
is satisfied for specific value of the potential
$V_0 (E) = V(E, x_0)$ in specific point $x_0$.
In the point $x = x_0 + \Delta x$, where $V(x_0 + \Delta x)$
is outside the $L_{coh}$-peak,
significant enhancement of coherence is absent and $L_{coh}^m < \Delta x$.

Certain increase of the coherence length can be related
to the fact that at $V_0$ the derivative
$d \Delta H_m(x)/d E$
changes the sign which suppresses the integral over $x$.
This corresponds to the change of sign of
the difference of the group velocities at $E_0$ ($V_0$).

For a layer with a given length $L$ the energy of zero
separation (complete overlap) is realized
(\ref{adiab-sep}) when
\begin{equation} \label{L_0}
 \int_0^{x_0} dx \frac{d}{d E} 
\Delta H_m(x)=-\int_{x_0}^{L} dx \frac{d}{d E} 
\Delta H_m(x) ,
\end{equation}
and $x_0 = x_0(E)$.

The infinite coherence can be achieved
in the potential with periodic modulations:
\begin{equation}
\label{eq:var-pot}
V(x) = \Bar{V} (1 + h \sin 2\pi x/X).
\end{equation}
For a large $X$, the adiabaticity condition is satisfied.
Zero separation of WP, $\Delta x = 0$, is realized when
the difference of group velocities has different signs in the
first and the second parts of the period.
So, there is a continuous "catch-up" effect.

According to (\ref{eq:partder}) the energy $E_0$ of infinite coherence
is obtained from the condition
\begin{equation}
\label{eq:intcond}
lim_{L \rightarrow \infty} \int_0^L dx
\frac{\left(\frac{\Delta m^2}{2 E} - V\cos 2\theta \right)}{\Delta
H_m(x)}
= 0.
\end{equation}
Around the zero value of nominator, denominator changes
much slower
and it can be put out of integral at some average value of the potential.
Then the condition (\ref{eq:intcond}) reduces to
\begin{equation}\nonumber
\label{eq:intcond2}
\frac{\Delta m^2}{2 E} L - \cos 2\theta \int_0^L dx V (x) = 0.
\end{equation}
Performing explicit integration with the potential
(\ref{eq:var-pot}) we obtain
\begin{equation}
\label{eq:intcond3}
\frac{\Delta m^2}{2 E_0} \approx \cos 2\theta \Bar{V} + \frac{h}{\pi}
\frac{X}{L}
\sin^2 \frac{\pi L}{X}.
\end{equation}
In the limit $L \rightarrow \infty$ the last term
on the RHS of (\ref{eq:intcond3}) can be neglected and for $E_0$
we obtain the same expression (\ref{eq-E0}) as in constant density case
with the average potential $\Bar{V}$.

In general, if $V$ varies about some average value $\bar V$ (even irregularly), there may be energy where $L_{coh}^m \rightarrow \infty$.

\subsection{Coherence for massless neutrinos}


The limit $L_{coh}^m \rightarrow \infty$ is realized 
for oscillations of massless neutrinos, 
as originally introduced by Wolfenstein \cite{Wolfenstein:1977ue}. 
Here the oscillation phase is independent of neutrino energy.
Recall that the propagation decoherence is related to 
the energy dependence of level splitting $\Delta H_m$ and 
shows up as averaging over energy. 
Therefore the decoherence does not exist for oscillations 
of massless neutrinos driven by potentials.

\section{Coherence and adiabaticity violation}
\label{coherence and adiabaticity violation}

Adiabaticity violation means transitions between the
eigenstates of propagation $\nu_{1m} \leftrightarrow \nu_{2m}$ and therefore change of shape of the WPs. Here we will consider the cases of extreme adiabaticity violation when $dV/dx \rightarrow \infty$, which corresponds to density jumps at certain spatial points.

Extreme adiabaticity violation combined with shift and separation of the WPs leads to splits and catch up of the WPs \cite{Kersten:2015kio}. We will first illustrate these effects considering a single density jump.

\subsection{Coherence in the case of single density jump}

Consider a two layers ($a$ and $b$) 
profile with density jump at a border between them. The layers have matter potentials $V_a$ and $V_b$ and lengths $L_a$ and $L_b$, 
so that the total length equals $L \equiv L_a + L_b$. 
Let $\theta_k = \theta_m(V_k)$
be the flavor mixing angle 
in layer $k$, 
($k = a, \, b$).

Suppose in the beginning of layer $a$ the state is
\begin{equation} 
\label{init}
\nu (x=0,t) = c_a' f_1(t) \nu_1^a + s_a' f_2 (t) \nu_2^a, 
\end{equation}
where we use abbreviations $c_a' \equiv \cos \theta_a$, 
$s_a' \equiv \sin \theta_a$ and 
$f_i(t)$ are the shape factors of the wave packets 
normalized as 
$$
\int dt |f_i (t)|^2 = 1.
$$
Evolving to the border between the layers $a$ and $b$ 
the state (\ref{init}) becomes 
\begin{equation} 
\label{border}
\nu (L_a,t) = c_a' f(t - t_1^a) e^{i 2\phi_1^a} \nu_1^a + 
s_a' f(t - t_2^a) e^{i 2\phi_2^a} \nu_2^a. 
\end{equation} 
Here $t_i^a$ are the times of propagation in the $a$-layer,
$t_i^a = L_a/ v_i^a$, $v_i^a$ are group velocities, and $\phi_i^a$ are phases 
acquired by the eigenstates taken 
at the average energies in the packets. 
We assume that in the beginning ($t=0$) the shape factors of $\nu_1^a$ 
and $\nu_2^a$ are equal. 

Crossing the border between $a$ and $b$ each eigenstate $\nu_i^a$ splits into eigenstates $\nu_j^b$ of the layer $b$:
\begin{equation} 
\label{splitab} 
\nu_1^a = c_\Delta \nu_1^b + s_\Delta \nu_2^b, \, \, \, \, 
\nu_2^a = c_\Delta \nu_2^b - s_\Delta \nu_1^b,
\end{equation}
where $\Delta \equiv \theta_b - \theta_a$. Inserting (\ref{splitab}) into (\ref{border}) we obtain the state in the beginning of layer $b$. Then to the end of the layer $b$ the state (\ref{border}) evolves to 
\begin{eqnarray} 
\label{border-ba}
\nu (L,t) = 
\left[ c_a' c_\Delta f(t - t_1^a - t_1^b) e^{i 2(\phi_1^a + 
\phi_1^b)} 
- s_a' s_\Delta f(t - t_2^a - t_1^b) e^{i 2(\phi_2^a + \phi_1^b)} 
\right] \nu_1^b + 
\nonumber \\
\left[c_a' s_\Delta f(t - t_1^a - t_2^b) 
e^{i 2(\phi_1^a + \phi_2^b)} 
+ s_a' c_\Delta f(t - t_2^a - t_2^b) e^{i 2(\phi_2^a 
+ \phi_2^b)} \right] \nu_2^b. 
\end{eqnarray} 
We put out the common phase factor 
$e^{i 2(\phi_1^a + \phi_1^b)}$ 
and introduce the phase differences,
\begin{equation} \label{phase-redefinition}
 \phi^a \equiv \phi_2^a - \phi_1^a \, \, \, \, \text{and} \, \, \, \, \phi^b \equiv \phi_2^b - \phi_1^b.
\end{equation}
Then we make the time shift as $t' = t - t_1^a - t_1^b$, which allows us to express results in terms of the relative 
time shifts of the eigenstates 1 and 2 in the layers $a$ and $b$:
\begin{equation} \label{time-redefinition}
 t^a \equiv t_2^a - t_1^a \, \, \, \, \text{and} \, \, \, \, t^b \equiv t_2^b - t_1^b.
\end{equation}


Using these quantities and projecting (\ref{border-ba}) 
onto $\nu_e$ we obtain the amplitude 
of $\nu_e \rightarrow \nu_e$ transition
over the two layer profile:
\begin{eqnarray} 
\label{eeamp1}
A_{ee} & = &
c_\Delta \left[
 c_a'c_b' f(t') e^{- i\phi}
+ s_a's_b' f(t' - t^a - t^b) e^{i\phi} 
\right] 
\nonumber \\
& + & 
s_\Delta \left[
 c_a' s_b' f(t' - t^b) e^{-i\phi'}
- s_a'c_b'f(t' - t^a) e^{i\phi'} 
\right]. 
\end{eqnarray} 
In this expression we put out the factor 
$e^{i(\phi^a + \phi^b)}$ and introduce
\begin{equation} \label{phase-redefinition-2}
 \phi \equiv \phi^a + \phi^b \, \, \, \, \text{and} \, \, \, \, \phi' \equiv \phi^a - \phi^b.
\end{equation}
\subsection{Specific examples}
Let us consider specific cases. 

1. Suppose $|t^a| \gg \sigma_t$, that is, 
the separation of the WPs in the layer $a$ 
is larger than the coherence time, which means that the 
WPs are completely separated and coherence is lost. Suppose also that 
$|t^b| \ll \sigma_t$, {\it i.e.} loss 
of coherence in the layer $b$ is negligible, $|t^b| \approx 0$. Then taking into account that the 
shape factors $f(t')$ and $f(t - t^a)$ do not overlap 
we obtain from (\ref{eeamp1}) 
\begin{eqnarray}
\label{eeamp2}
P_{ee} & = & \int dt |A_{ee}|^2 = 
\int dt |f(t')|^2 
\left|c_a' c_b' c_\Delta + 
c_a' s_b' s_\Delta e^{ i 2\phi^b} \right|^2
\nonumber \\
& + & \int dt |f(t' - t^a)|^2 
\left| s_a's_b' c_\Delta e^{i 2\phi^b} 
- s_a'c_b's_\Delta \right|^2. 
\end{eqnarray} 
We assume that change of the oscillation phase $\phi^b (t)$ along the 
WP is negligible, so that the oscillatory factors can be
put out of the integral and take into account that 
integrations of the moduli squared of the shape factors give $1$ 
due to normalization. As a result, we obtain 
\begin{equation}
\label{eeamp2}
P_{ee} = 
\left|c_a' c_b' c_\Delta + 
c_a' s_b' s_\Delta e^{ i 2\phi^b} \right|^2
+ \left| s_a's_b' c_\Delta e^{i 2\phi^b} 
- s_a'c_b's_\Delta \right|^2. 
\end{equation} 

Computing the amplitudes in the energy space, we need to use the plane waves so that $f(t) =1 $, and the energy
dependent phases $\phi^a (E)$ and $\phi^b (E)$. 
The probability is given by 
\begin{equation}
\label{eepr1}
P_{ee} = \int dE F(E) |A_{ee}|^2,
\end{equation}
where $F(E)$ is the energy spectrum of neutrinos 
that corresponds to the WP in the $E$-$p$ space
$F(E) = |f(E)|^2$ and $f(E)$ is the Fourier transform of $f(t)$. 
The amplitude (\ref{eeamp1}) can be written as 
\begin{equation} 
\label{eeamp12}
A_{ee} = e^{-i\phi(E)} A_1 + e^{i\phi'(E)} A_2, 
\end{equation}
with
\begin{equation} 
\label{a1a2}
A_1 = c_a' \left[c_b' c_\Delta 
+ s_b' s_\Delta e^{i2\phi^b(E)} \right], 
\, \, \, \
A_2 = s_a'\left[s_b'c_\Delta e^{i 2\phi^b(E)} 
- c_b' s_\Delta \right]. 
\end{equation} 
Here $A_i$ are the oscillations amplitudes of the eigenstates 
of the layer $a$, $\nu^a_i$, in the layer $b$, $\nu_i^a \rightarrow \nu_e$. 
Inserting (\ref{eeamp12}) into (\ref{eepr1}) we have 
\begin{equation}
\label{eepr3}
P_{ee} = \int dE F(E) \left[ |A_1|^2 + |A_2|^2 + 
2|A_1 A_2| \cos 2(\phi^a + \chi) \right], 
\end{equation}
where $\chi \equiv Arg (A_1^* A_2)$ depends on the phase $\phi^b$. 
To compare this with consideration in the $x$-space we assume that 
$\phi^a \gg 1$ (which corresponds to large delay in the layer $a$). In contrast, 
the phase $\phi^b$ is relatively small, 
$\phi^b \ll 1$. In this case we can put all the terms 
but those which depend on $\phi^a$ 
in (\ref{eepr3}) out of the integral taking them
at average value of $E$ in the spectrum. Then 
due to normalization $\int dE F(E) = 1$ the expression 
(\ref{eepr3}) becomes
\begin{equation}
\label{eepr4}
P_{ee} = |\bar{A}_1|^2 + |\bar{A}_2|^2 + 
2|\bar{A}_1 \bar{A}_2| \int dE F(E)\cos 2(\phi^a + \chi). 
\end{equation}
Here $\bar{A}_i$ are the amplitudes at the average value of energy. 
The last term is suppressed as $1/\phi^a$ and negligible 
for $\phi^a \gg 1$. Thus, the result (\ref{eepr4}), with $A_i$ in (\ref{a1a2}), coincides with that in the $x$-representation (\ref{eeamp2}). 

In the example considered above, the result does not depend
on $L_a$, and $L_a \rightarrow \infty$ is possible. In a sense,
one can consider this as the case of infinite coherence since
oscillations
can be observed at the arbitrary long distance from the source.
Thus, density jump and splitting of eigenstates induce or restore
the interference and therefore oscillations. Here suppression of
propagation
coherence is determined
by properties of layer $b$. That is, the problem is reduced to
a single layer with constant density.
Depending on initial mixing and $\Delta \theta$, crossing
the density jump can lead to even stronger
interference than at the beginning of layer $a$.

The situation described above is realized, {\it e.g.}, 
for the solar and supernova neutrinos oscillating inside the Earth, 
where $a$ is the vacuum between the star and the Earth, and $b$ is the Earth. 
In the high energy part of solar neutrinos: $c_a^2 \ll s_a^2$.

2. Suppose the layer $b$ has density at which
velocities of eigenstates are equal so that coherence is maintained for arbitrary large $L_b$ and $t^b=0$.
Furthermore, suppose in the layer $a$ the wave packets
are completely separated, $L_a \rightarrow \infty$ which means that dependence on $\phi^a$ disappears.
Then according to (\ref{eeamp1}),
\begin{eqnarray}
\label{eeampcomp}
A_{ee} & = &
c_a' c_b' c_\Delta f(t) e^{i \phi} - s_a'c_b' s_\Delta f(t - t^a) e^{i\phi'}+
\nonumber \\
& & c_a' s_b' s_\Delta f(t) e^{-i\phi'}
+ s_a's_b' c_\Delta f(t - t^a) e^{i\phi}.
\end{eqnarray}
The interference of the overlapping parts
(they have the same argument in $f$) equals
\begin{equation}
\label{int1112}
t: \, \, 2 c_a'^2 c_b' s_b' s_\Delta c_\Delta \cos 2\phi^b (t),
\, \, \, \, \, \,
(t - t^a): \, \, - 2 s_a'^2 c_b' s_b' s_\Delta c_\Delta
\cos 2\phi^b (t - t^a).
\end{equation}
At a detector, both oscillation phases
are equal to $\phi^b$, and, the sum of interference terms (after integration over
time)
is
\begin{equation}
\label{intsum}
\cos 2\theta_a \sin 2 \theta_b s_\Delta c_\Delta \cos 2\phi^b.
\end{equation}
This can be compared to the depth of interference
in the beginning of the layer $a$: $0.5\sin^2 2\theta_a$.

Regions of parameters exist where the depth increases after
propagation in the two-layer profile. \\

Apart from the restoration of the interference, there is another phenomenon
which will be important for our further consideration, namely, change of the shape of the WP at
crossing the border between the layers:

\begin{itemize}

\item
Each packet $\nu_i^a$ becomes the two-component one with an overall size
extended by the relative delay in the layer $a$, $t^a$.

\item
According to (\ref{eeampcomp})
the amplitudes of components of $\nu_1^b$ equal
$(c_a' c_\Delta, \, \, - s_a' s_\Delta)$ for the earlier and the later
ones correspondingly. The amplitudes in $\nu_2^b$ are
$(c_a' s_\Delta, \, \, s_a' c_\Delta)$.

\item
The shape depends on the size of density jump.
If, {\it e.g.}, the jump is small, $s_\Delta \sim s_a' \sim \epsilon \ll 1$,
we find from the previous item the amplitudes
for $\nu_1^b$: $(1, \, \epsilon^2)$,
and for $\nu_2^b$: $(\epsilon, \, \epsilon)$.
If $s_\Delta \approx c_\Delta$, the packets will have similar form with similar amplitudes
$\nu_1^b$: $(c_a', \, - s_a')$ and $\nu_2^b$: $(c_a', \, s_a')$.
If $s_\Delta = 1$, $c_\Delta = \epsilon$ we have
$\nu_1^b$: $(\epsilon, \, \epsilon)$ and
$\nu_2^b$: $(1, \, \epsilon^2)$.

\end{itemize}

3. Consider the case $t^a = - t^b$, when separation in the layer $b$
compensates separation in $a$. In spite of this compensation,
and in contrast to the constant density or adiabatic cases,
there is no complete overlap at the end of layer $b$
due to split of the eigenstates. Only WP of $A_{11}$ and $A_{22}$
components will overlap completely, where subscripts indicate eigenstates in which a given final component propagated in layer $a$ and $b$ ({\it e.g.} $A_{11}$ is the amplitude of WP $\nu_1^a \rightarrow \nu_1^b$). The component $A_{12}$
will shift forward by $t^a$ and $A_{21}$ backward by $t^a$ with respect
to
overlapping components,
or vice versa, depending on the sign of $t^a$.
If $|t^a| > \sigma_x$, only $A_{11}$ and $A_{22}$ interfere and
the interference term equals
\begin{equation}
\label{intdd}
0.5 \sin 2\theta_a \sin 2\theta_b c_\Delta^2 \cos 2(\phi^a + \phi^b),
\end{equation}
which again can be bigger than interference in
the beginning of the layer $a$. \\

4. In the case $t^a \approx 0$, there is no separation and loss
of coherence in the layer $a$. The components $A_{11}$ and $A_{21}$
as well as $A_{22}$ and $A_{12}$ will overlap, there is no
split of eigenstates at the border $a \rightarrow b$.

There is no unambiguous way of introducing the coherence length in the presence of density jump in the profile. Recall that the equivalence in the $E$- and $x$-representations 
is established at the level of averaging of the oscillation phase and separation of the WP. 
In the case of a single jump, there are two different oscillation phases and two different delays due to splitting of eigenstates. The introduction of a single effective phase and effective delay is non-trivial. 
As we will see in sect.~\ref{parametric-oscillations}, this can be done for periodic structures with density jumps.

\subsection{Wave packets in the $x$- and $E$-representations}

Split of the eigenstates and delays of splitted components lead
to substantial modifications of the shape factor
of the total WPs, $\psi_i(x,t)$, in the $x$-space. 
At the same time, the Fourier transforms of $\psi_i(x,t)$
determine the neutrino energy spectrum. Let us show that despite substantial modifications
of $\psi_i(x,t)$, the energy spectrum remains unchanged, as it should be \cite{Stodolsky:1998tc}.

Suppose $f(E)$ is the Fourier transform of $f(t)$,
then in the $E$-representation the initial state (\ref{init}) is
$$
\nu(x=0,E) = f(E)(c_a' \nu_1^a + s_a' \nu_2^a).
$$
The energy spectrum of this state equals
\begin{equation}
F(E) = |f(E) c_a'|^2 + |f(E) s_a'|^2 = |f(E)|^2.
\label{eq:spectr1}
\end{equation}

Consider the simplest element of the profile
which leads to modification of the shape factor:
the layer $a$ with constant density and
density jump at the end
(general case is just repetition of this element). State (\ref{border-ba}), at the beginning of layer $b$ ($t^b_i = 0$ and $\phi^b_i = 0$) in terms of $\psi_i(x=L_a,t)$ is
\begin{equation}
\nu(L_a,t) = \psi_1(L_a,t) \nu_1^b + \psi_2(L_a,t) \nu_2^b,
\label{total-WPs}
\end{equation}
where the total WPs equal
\begin{eqnarray} \label{total-WPs-2}
\psi_1(L_a,t) & = &
[c_a' c_\Delta f(t) - s_a' s_\Delta f(t - t^a) e^{2i \phi^a}],
\nonumber\\
\psi_2(L_a,t) & = &
[c_a' s_\Delta f(t) + s_a' c_\Delta f(t - t^a) e^{2i \phi^a}].
\label{eq:totwp12}
\end{eqnarray}
Here we made time shift by $t_1^a$ and put out common phase
$\phi^a_1$.
In the beginning, $\psi_i$ are the elementary WPs, $f(t) \nu_i^a$, as in (\ref{init}), but along the propagation their widths increase and their forms change.

In $E$-space, the state just after crossing the density jump, $\nu(L_a)$, in Eq.~(\ref{total-WPs}), is:
\begin{equation}
\nu(L_a,E) = \psi_1(L_a,E) \nu_1^b + \psi_2(L_a,E) \nu_2^b,
\label{eq:statexb}
\end{equation}
The Fourier transform of
$f(t - t^a)$ is $f(E) e^{i E t^a}$, so that
\begin{eqnarray}
\psi_1(L_a,E) & = &
[c_a' c_\Delta - s_a' s_\Delta e^{- i E t^a + 2i \phi^a}] f(E) ,
\nonumber\\
\psi_2(L_a,E) & = &
[c_a' s_\Delta + s_a' c_\Delta e^{- iEt^a + 2i \phi^a}] f(E).
\nonumber
\end{eqnarray}
Using these expressions, we find the energy spectrum
$$
F(E) = |\psi_1(L_a,E)|^2 + |\psi_2(L_a,E)|^2 = |f(E)|^2,
$$
which coincides with (\ref{eq:spectr1}).

\section{Coherence in the castle-wall profile}
\label{parametric-oscillations}

One can introduce the coherence length in the case of a periodic structure with a sharp density change. The castle-wall (CW) profile 
is multiple repetitions of the two layer structure considered sect.~\ref{coherence and adiabaticity violation}. 
This is an explicitly solvable example 
of the periodic profile, which can give an idea about effects in 
profiles with sine (or cosine) type of dependence on distance \cite{Akhmedov:1988kd,Akhmedov:1999ty,Akhmedov:1999va}. 

\subsection{Parametric oscillations in the energy-momentum space} \label{parametric1}

The oscillation probability after crossing $n$ periods can be obtained from 
the probability for a single period computed above. Indeed, the amplitude $A_{ee}$ (\ref{eeamp12}),(\ref{a1a2}) can be written as 
\begin{equation} \nonumber
\label{eeprob}
A_{ee} = R + i I_3, 
\end{equation}
where the real and imaginary parts of the amplitude equal respectively, 
see \cite{Akhmedov:1999ty,Akhmedov:1999va}, 
\begin{equation} 
\label{R}
R=c_{a} c_{b} - s_{a} s_{b} \cos \left(2 \theta_a - 2 \theta_b\right),
\end{equation} 
and 
\begin{equation} \nonumber
\label{third-component}
I_3 = - (s_a c_b \cos 2 \theta_a + s_b c_a \cos 2 \theta_b).
\end{equation}
Then the probability is given by
\begin{equation} \nonumber
\label{eeprob}
P_{ee} = R^2 + I_3^2. 
\end{equation}

Similarly to $A_{ee}$ one can find the amplitudes 
$A_{e \mu}$, $A_{\mu e}$ and $A_{\mu \mu}$ which compose 
the evolution matrix $U_L$ over one period. The matrix 
allows to reconstruct the Hamiltonian integrated over the period: 
$i H_{CW} L = I - U_L$. Diagonalization of this Hamiltonian 
gives the eigenstates, {\it i.e.} mixing and 
difference of eigenvalues 
which allow to write the transition probability 
after passing $n$ periods of size $L$, $P = |U(nL)_{\alpha \beta}|^2$, as 
\begin{equation} 
\label{P-CW}
P(\nu_{\alpha} \rightarrow \nu_{\beta},nL) = 
\left(1-\frac{I_3^2}{1 - R^2} \right) \sin^2 \left(n \xi \right).
\end{equation}
Here $\xi$ is the effective phase acquired over period:
\begin{equation}\nonumber
\xi \equiv \arccos R. 
\end{equation}
The factor in front of sine gives the depth of the parametric 
oscillations. The depth is maximal at $I_3 = 0$, or explicitly
\begin{equation} 
\label{condition1}
 s_a c_b \cos 2 \theta_a +c_a s_b \cos 2 \theta_b = 0, 
\end{equation}
which is the condition of the parametric resonance. 
 For arbitrary mixings the condition is satisfied 
for values of phases
\begin{equation} \nonumber
\label{condition2}
\phi^a = \frac{\pi}{2}+k \pi \hspace{0.5 cm} \text{and} 
\hspace{0.5 cm} 
\phi^b = \frac{\pi}{2}+k' \pi. 
\end{equation}
In general, $I_3 = 0$ requires specific correlations between 
phases and mixings. The transition probability (\ref{P-CW}) reaches maximum, $P = 1$, when 
\begin{equation} \nonumber
\label{condition3}
n \xi = \frac{\pi}{2} + k \pi.
\end{equation}

In what follows for illustration of the results we will use 
a castle-wall profile with $n = 5$ periods 
and the following parameters: 
\begin{equation} 
\label{eq:paramet}
V_a = 5.3 \times 10^{-4} \, \text{eV}^{2}, \, \, \, \, \, L_a = 4 \, {\rm km}, \, \, \, 
V_b = 2 \times 10^{-5} \, \text{eV}^{2}, \, \, \, L_b = 2 \, {\rm km}.
\end{equation}
We take the vacuum oscillation parameters $\theta=8.5^{\circ}$ and 
$\Delta m^2=2.5 \times 10^{-3}$ $\text{eV}^2$. 

\begin{figure} 
\centering
\includegraphics[width=0.7\linewidth]{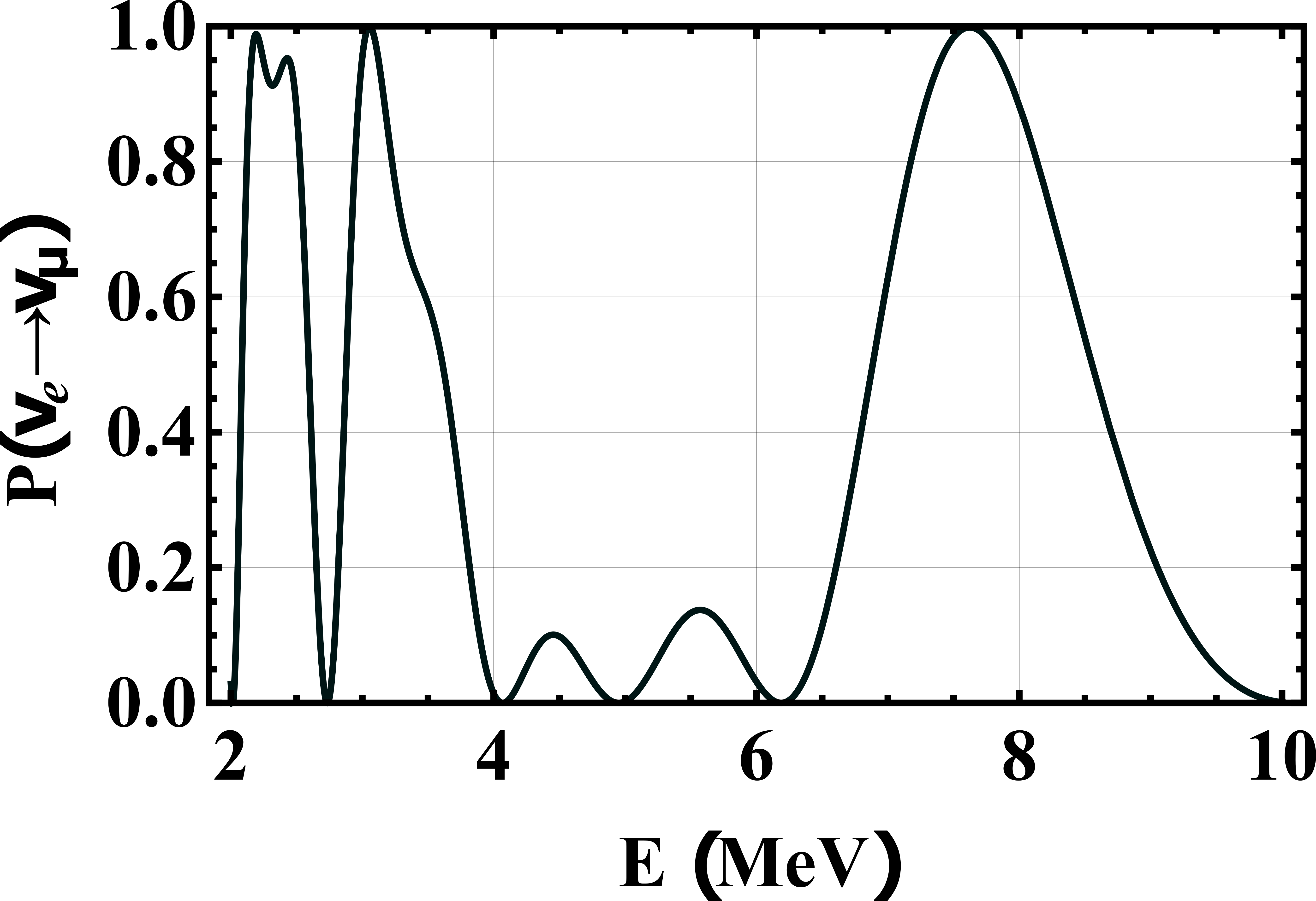}
\caption{
Transition probability of the $\nu_e \rightarrow \nu_{\mu}$ parametric oscillations as a function of neutrino energy for castle-wall profile with $5$ periods. 
Parameters of the CW profile are given in Eq.~(\ref{eq:paramet}).
We used $\theta=8.5^{\circ}$ and $\Delta m^2=2.5 \times 10^{-3}$ $\text{eV}^2$.}
\label{resonance1}
\end{figure}

The transition probability of parametric oscillations 
Eq.~(\ref{P-CW}), as function of energy, is shown in fig.~\ref{resonance1}. The probability
reaches maximum at $E \sim 2.2$ (MSW resonance), $3.1$ and $7.7$ MeV, 
where condition (\ref{condition1}) is met.

\subsection{Coherence length in a castle-wall profile}
\label{coherence-castle-wall}

In the energy representation we can repeat here the same procedure of determination of the 
coherence length as in the case of constant density. 
According to Eq.~(\ref{P-CW}), the phase of parametric 
oscillations equals
\begin{equation} \nonumber
\label{CW-phase}
 \phi_n = n \xi = n L \frac{\xi}{L} . 
\end{equation}
Therefore the effective difference of eigenvalues equals 
\begin{equation} \nonumber
\label{eff-eig}
\Delta H_{cw} = \frac{\xi}{L}. 
\end{equation}

We introduce $n_{coh}$, the number of periods over which the coherence 
is maintained, so that $L_{coh}=n_{coh} L$. 
Using Eq.~(\ref{const-density}) 
for the coherence length we find
\begin{equation} \nonumber
\label{phase-var}
n_{coh} = \frac{\pi}{\sigma_E}
\left|\frac{d \xi}{d E} \right|^{-1}. 
\end{equation}
According to (\ref{const-density}) and (\ref{R})
\begin{equation} 
\label{D-xi}
\frac{d \xi}{d E} = 
-\frac{1}{\sqrt{1-R^2}} \frac{d R}{d E}, 
\end{equation}
and consequently, 
\begin{equation} 
\label{p-coh}
n_{coh}(E)=\frac{\pi \sqrt{1-R^2}}{2 \sigma_E \left| 
\frac{d R}{d E} \right|}.
\end{equation}

The condition for infinite coherence length, 
$L_{coh} \propto n_{coh} \rightarrow \infty$, is 
\begin{equation}
\label{divergency} 
\left|\frac{d \xi}{d E} \right| = 0, \, \, \, \, 
{\rm or } \, \, 
\frac{d R(E)}{d E} = 0, \, \, \, (R \neq 1).
\end{equation}
Since $R(E)$ is an oscillatory function of $E$, and therefore has several 
maxima and minima at certain energies $E_0^i$, the condition 
(\ref{divergency}) is satisfied at these energies 
$E = E_0^i$.

As in the case of constant density, averaging of probability over the 
energy intervals $\sigma_E$ leads to the appearance 
of the decoherence factors 
$D_i(E)$, (\ref{decoh}), centered at energies $E_0^i$. Let us find the widths 
of the peaks of $D_i(E)$ for fixed length of trajectory 
$ n L$. 
As in constant density case, instead of averaging over 
$E$ we perform averaging over 
the effective phase in the interval which corresponds to 
$2\sigma_E$: 
\begin{equation}
\label{eq:dphi-n}
\delta \phi_n = 2\sigma_E n \frac{d \xi}{d E}.
\end{equation}
The decoherence factor 
\begin{equation}\nonumber
 D(\delta \phi_n (E)) = \frac{\sin \delta \phi_n}{\delta \phi_n}
\end{equation}
has a width $\Gamma$ determined by the condition in (\ref{Gamma}).
Plugging 
$\delta \phi_n$ from (\ref{eq:dphi-n}) in the above equation we obtain
\begin{equation} 
\label{half-max1}
\left| \frac{d \xi(E_0^i +\Gamma)}{d E} \right| 
= \frac{1}{\sigma_E n}.
\end{equation}
For narrow peaks, $\Gamma \ll E_0^i$, 
we can expand the left hand side of Eq.~(\ref{half-max1}) 
around $E_0$; and take 
into account that 
$d \xi (E_0)/d E = 0$. Then Eq.~(\ref{half-max1})
reduces to 
\begin{equation}
\frac{d^2 \xi (E)}{d E^2}\bigg{|}_{E=E_0^i} 
\times \Gamma = \frac{1}{\sigma_E n}.
\label{eq:condigamma}
\end{equation}
Differentiating $d\xi/dE$ in (\ref{D-xi}) over $E$ we have 
\begin{equation}
\label{eq:doubleder}
\frac{d^2 \xi (E)}{d E^2}\bigg{|}_{E=E_0^i} = 
-\frac{1}{\sqrt{1- R^2(E_0^i)}} \frac{d^2 R(E_0^i)}{d E^2}.
\end{equation}
Then insertion of this into (\ref{eq:condigamma}) gives
\begin{equation} \nonumber
\label{CW-peak}
\Gamma \approx 
\frac{\sqrt{1-R^2(E_0^i)}}{ \sigma_E n \left| 
\frac{d^2 R(E_0^i)}{d E^2} \right|}. 
\end{equation}

In the limits $E_0^i \ll E_{a},E_{b}$ and $E_0^i \gg E_{a},E_{b}$,
where $E_{a}$ and $E_{b}$ are the MSW resonance energies, we have 
$\theta_a \approx \theta_b$, and therefore 
\begin{equation} \nonumber
\label{approx-R}
R \approx \cos (\phi^a + \phi^b). 
\end{equation}
Since $d\xi/dE \propto d(\phi^a + \phi^b)/dE = 0$ at $E = E_0^i$ 
the equation (\ref{eq:doubleder}) becomes 
\begin{equation}\nonumber
\frac{d^2 \xi}{d E^2} \bigg{|}_{E=E_0} 
\approx \frac{d^2 (\phi^a + \phi^b)}{d E^2}\bigg{|}_{E=E_0},
\end{equation}
and the width equals
\begin{equation} \nonumber
\label{approx-gamma}
\Gamma_i \approx \frac{1}{\sigma_E n 
\frac{d^2 (\phi^a + \phi^b)}{d E^2} \big{|}_{E=E_0^i} }.
\end{equation}
The widths $\Gamma^i$ are narrower at lower energies 
due to faster variations of the half-phases $\phi^a$ 
and $\phi^b$. Similarly to the case of constant 
matter density, the ranges of weak averaging effect 
of transition probability are smaller at lower $E_0^i$. 

\subsection{Infinite coherence and parametric resonance}

The energies of infinite coherence, $E_0^i$, 
are correlated to the energies of parametric resonances, 
determined by the condition $I_3=0$. This correlation is analogous to the one in Eq.~(\ref{MSW}) for constant density.
To show this, we consider the expressions for $d R/d E$ and $I_3$ 
in three different regions of energies. Starting with the explicit expression for $d R/d E$,
\begin{multline} \label{dR}
\frac{d R}{d E}=-s_a c_b \frac{d \phi^a}{d E}-c_a s_b \frac{d \phi^b}{d E} -
c_a s_b \cos(2\theta_a-2 \theta_b) \frac{d \phi^a}{d E}\\
-s_a c_b \cos(2\theta_a-2 \theta_b) \frac{d \phi^b}{d E} + 
s_a s_b \sin(2\theta_a-2 \theta_b) \frac{d (2\theta_a-2 \theta_b)}{d E}.
\end{multline}
We obtain the following
\begin{enumerate}

\item Far from the MSW resonances of both layers:
$E \ll E_{a}, \, E_{b}$, 
where $\theta_a \approx \theta_b \approx \theta$, or 
$E \gg E_{a}, \, E_{b}$ where 
$\theta_a \approx \theta_b \approx \pi/2$. 
In both cases $\cos (2\theta_a - 2\theta_b) \approx 1$,
and consequently, $R \approx \cos (\phi^a + \phi^b)$. Therefore 
\begin{equation} \nonumber
 \frac{d R}{d E} = 
- \sin (\phi^a + \phi^b)\frac{d (\phi^a + \phi^b) }{d E} 
\end{equation}
On the other hand 
\begin{equation}\nonumber
I_3 \approx \sin (\phi^a + \phi^b) \cos 2\theta_a.
\end{equation}
So that, the condition $I_3 \rightarrow 0$ requires 
$\sin (\phi^a + \phi^b) = 0$ and consequently, 
$\frac{d R}{d E} \rightarrow 0$, 
which implies $n_{coh} \rightarrow \infty$.

\item In the MSW resonances: when $E$ coincides with one of the resonance 
energies {\it e.g.} $E \approx E_{a}$, 
and is far enough from the other one, $E_{b}$, then
$\theta_a=\pi/4$, $\theta_b = \theta \approx 0$ (small mixing) or $\pi/2$ and $d \theta_b/d E=0$. Because $E_{a}\sim E_{0a}$, we can take in (\ref{dR}) $d \phi^a/d E=0$ which becomes
\begin{equation} \label{dR1}
 \frac{d R}{d E}=(-c_a s_b \pm s_a c_b \cos 2\theta_a) \frac{d \phi^b}{d E}\pm 2s_a s_b \sin 2\theta_a \frac{d \theta_a}{d E}.
\end{equation}
Here the upper sign is for $\theta_b \approx 0$ and the lower one $\theta_b \approx \pi/2$. We can rewrite (\ref{dR1}) as
\begin{equation} \label{dR2}
 \frac{d R}{d E}=\pm I_3 \frac{d \phi^b}{d E}\pm 2s_a s_b \sin 2\theta_a \frac{d \theta_a}{d E}.
\end{equation}
Assuming that $\sigma_E d \phi^b/d E \sim 1$ and $\sigma_E d \theta_a/d E \ll 1$, we find that the second term in (\ref{dR2}) shifts the zero of $d R/d E$ relative to the zero of $I_3$. Nevertheless, $E_0$ is close to $E_R$, and the closer they are, the smaller $d \theta_a/d E$ is relative to $d \phi^b/d E$.

\item Between the MSW resonances:
$E_{a} < E< E_{b}$ or $E_{a} > E > E_{b}$. 
If resonances are well separated, 
we have $|\theta_a - \theta_b| \approx \frac{\pi}{2}$ or 
$\cos (2\theta_a - 2\theta_b) \approx -1$. Therefore, 
\begin{equation}\nonumber
R \approx \cos (\phi^a - \phi^b), 
\end{equation}
and 
\begin{equation}\nonumber
 \frac{d R}{d E} = 
- \sin (\phi^a - \phi^b)\frac{d (\phi^a - \phi^b) }{d E}. 
 \end{equation}
On the other hand, 
\begin{equation}\nonumber
I_3 \approx -\sin (\phi^a - \phi^b) \cos 2 \theta_1.
\end{equation}
Consequently, $I_3 \rightarrow 0$ implies 
$d R/ d E \rightarrow 0$, and 
$n_{coh} \rightarrow \infty$. 

\end{enumerate}

\begin{figure} 
\centering
\includegraphics[width=1\linewidth]{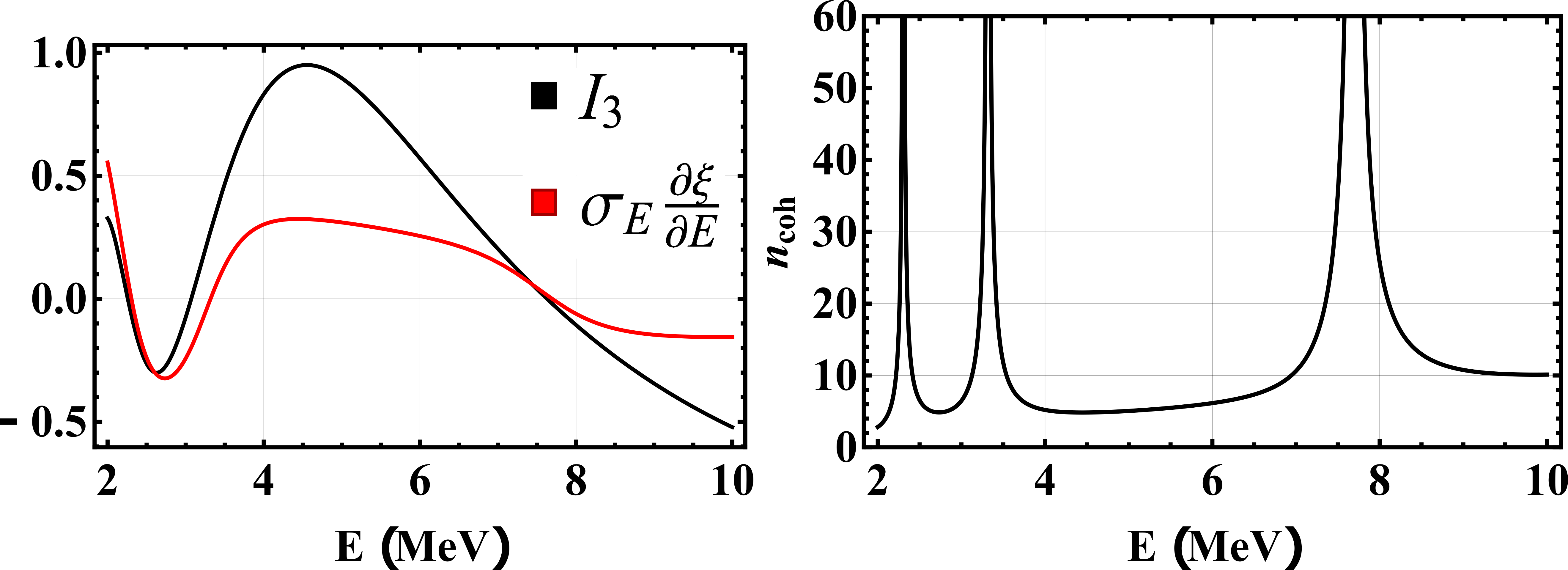}
\caption{Left panel: $I_3$ and $\frac{d \xi}{d E}$ as function 
of neutrino energy. Right panel: Dependence of the $n_{coh}$ in (\ref{p-coh}) 
on neutrino energy with $\sigma_E/E=0.1$ MeV. The parameters of CW profile are the same as in fig.~\ref{resonance1}. 
}
\label{fig-p-coh}
\end{figure}

In the left panel of fig.~\ref{fig-p-coh} we show $d R/d E$
and $I_3$ as functions of neutrino energy.
$I_3$ is null at $2.25$, $3.1$ and $ 7.7$ MeV, while the zeros of 
$d R/d E$, $E_0^i$ are $2.3$, $3.3$ and $7.7$ MeV. 
These coincidences show correlation of the parametric resonance and infinite coherence.
The MSW resonance energies are $E_{a}=2.25$ MeV 
and $E_{b}=62.7$ MeV, outside the range of the plot.

On the right panel of fig.~\ref{fig-p-coh} we show dependence of $n_{coh}$ on $E$
for the CW profile with parameters (\ref{eq:paramet}) 
and $\sigma_E/E=0.1$ MeV. $n_{coh}$ diverges at the energies $E_0^i$. Notice that the shape 
of the peak is similar to MSW case in fig.~\ref{fig-E0}. Thus, at the parametric resonance energies 
the averaging of oscillations is weak, and coherence is enhanced.

\subsection{Coherence in castle-wall 
profile in the configuration space}

In the $x$-representation, a picture of evolution in the CW profile is the following. 
Crossing every border between the layers $a$ and $b$ 
a given eigenstate of the layer $a$, $\nu_i^a$, splits into two eigenstates of the layer $b$: $\nu_i^a \rightarrow \nu_1^b, \nu_2^b$. In turn, at the next border each eigenstate $\nu_j^b$ splits into eigenstates 
of $a$ $\nu_j^b \rightarrow \nu_i^a$,
{\it etc}. The amplitudes of the splits are determined by 
change of the mixing angles in matter (\ref{splitab}). 

As in subsection~\ref{parametric1}, \ref{coherence-castle-wall} we consider that a flavor neutrino enters the layer $a$ of the first CW period and 
a detector is placed at the border of $b$ layer of the last period. After crossing $n$ periods of the CW profile, 
each initially produced eigenstate $\nu_i^a$ ($i = 1, 2$) splits 
$2^{2n -1}$ times, and therefore at a detector 
there are $2^{2n}$ elementary components. 
The eigenstate $\nu_i^b$ ($i=1,2$), that arrives 
at a detector (last border), will be a composition 
of $2^{2n -1}$ such components. They will arrive at different moments and have different phases and amplitudes. 
We will assume that these elementary WPs are short enough and do not spread. Therefore the oscillation phases (phase differences) are the same along the elementary WPs. Each elementary WP has its "history" of propagation determined by 
type of eigenstate it showed up in each layer of the profile, 
{\it e.g.} $\nu_1^a \rightarrow \nu_2^b \rightarrow \nu_2^a \rightarrow \nu_1^b \rightarrow \nu_1^b \rightarrow ... \rightarrow \nu_2^b$. 

The elementary WPs compose the total WP of eigenstates $\psi_i$.
Splits of eigenstates and delays lead to spread
of total WP after crossing the CW profile.
This spread affects the level of overlap
and coherence.
If a detector is sensitive to the flavor
$\nu_f = c_b' \nu^b_1 + s_b' \nu^b_2$, the detected signal is determined by 
\begin{equation}
\label{eq:sign}
\int dt \left| c_b' \psi^b_1 (t) + s_b' \psi^b_2 (t) \right|^2. 
\end{equation}
where $\psi_i^b$ are the total WP of the eigenstates
$\nu_i^b$ in a layer $b$ at a detector.
The total WP, $\psi^b_j (t)$, can be found summing up all $2^{2n-1}$ elementary WPs at the detector.

A given elementary WP at the detector can be characterized by 
$k_a$ and $k_b$ - the numbers of $a$ and $b$ layers it propagates through as 
the second eigenstate, {\it i.e.} $\nu_2^m$, $m=a,b$. 
Correspondingly, the numbers of layers 
it crossed as $\nu_{1}^m$ are $(n - k_a)$ and $(n - k_b)$. Using the same notation for phases, $\phi_i^{a}$ and $\phi_i^{b}$ $(i = 1,2)$, as in sect.~\ref{coherence and adiabaticity violation} $(i = 1,2)$ we can write 
the total phase of a given elementary packet 
at a detector: 
\begin{equation}
\label{eq:totph}
\phi (k_a, k_b) = \phi_1^{a} (n - k_a) + \phi_2^{a} k_a 
+ \phi_1^{b} (n - k_b) + \phi^{b}_2 k_b. 
\end{equation}
The phase difference between the elementary WPs characterized by 
$k_a, k_b$ and $k_a', k_b'$ equals
\begin{equation}
\label{eq:totphdif}
\phi (k_a, k_b, k_a', k_b') = 
\phi^{a} (k_a - k_a') + \phi^{b} (k_b - k_b'), 
\end{equation}
where we used definitions in (\ref{phase-redefinition}).

Similarly one can find the relative time delays of arrival of 
the packets characterized by $k_a, k_b$ and $k_a', k_b'$ at a detector using the definitions (\ref{time-redefinition}): 
\begin{equation}
\label{eq:totdeldif}
t (k_a, k_b, k_a', k_b') = t^a (k_a - k_a') + t^b (k_b - k_b'). 
\end{equation}
As for the case of a single layer (\ref{eq:phase-delay}), there are the relations between the 
total delays (\ref{eq:totdeldif}) and total oscillation phases (\ref{eq:totphdif}):
\begin{equation}
\label{eq:correltp}
t (k_a - k_a', k_b- k_b') = 
\frac{d \phi (k_a - k_a', k_b - k_b')}{d E}, 
\end{equation}
and 
\begin{equation}
\label{eq:correltp}
t (k_a - k_a', k_b- k_b') = 
g(E, V) \phi (k_a - k_a', k_b - k_b'). 
\end{equation}

Suppose, for definiteness, that $t^a$ and $t^b$ are positive, 
then maximal relative delay (separation) $t^{max}$ corresponds to 
$|k_a - k_a'| = |k_b - k_b'| = n$: 
$$
t^{max} = n(t^a + t^b).
$$
This is the time difference of arrival of the fastest 
and the slowest extreme elementary WP. 

$t^{max}$ separation determines the spread of whole WPs. The space between the fastest and the slowest WPs is 
filled in by non-extreme WPs. 
The extreme WPs are coherent if 
\begin{equation}
\label{eq:nxs-coh}
n (t^a + t^b) \leq \sigma_x,
\end{equation}
which gives the coherence number of periods (length) 
\begin{equation} 
\label{part-1}
n_{coh} \leq \frac{\sigma_x}{t^a + t^b}.
\end{equation}
Under this condition, all other (intermediate) WP-components 
are also coherent (overlap). 
The condition (\ref{part-1}) is sufficient but not necessary, since 
the coherence broken for extreme packets 
still can hold for intermediate packets. 

Parameters $k_a, k_b$ uniquely determine the phase and the delay of a given component. 
Number of components with a given 
$k_a$ and $k_b$ equals
$$
 ^n_{k_a}C \times~ ^n_{k_b}C, 
$$
where $^n _{p}C$ is number of combinations of $p$ elements from $n$ elements. Apart from 
$k_a$ and $k_b$
the amplitudes of individual components 
are determined by the product of various mixing parameters: the initial and final flavor mixing, {\it e.g.} $c_a'c_b'$ for $\nu_e \rightarrow \nu_e$ channel and $s_\Delta$, $c_\Delta$ determined by change 
of mixing at the borders between the 
layers $\Delta$:
$$ 
A(n,r,h) = (-1)^h c'_a c'_b s_\Delta^r c_\Delta^{2n - 1 - r}. 
$$
Factor $s_\Delta$ originates from each border at which the eigenstate number 
changes: $\nu_{1}^a \rightarrow \nu_{2}^b$, \textit{etc.}, while $c_\Delta$ 
appears from the borders without change of eigenstate: 
$\nu_{i}^a \rightarrow \nu_{i}^b$. $r$ is the number of borders where eigenstate changes; it is 
determined by the number of merging blocks: sequences of layers without change 
of eigenstate. $h$ is given by the number of blocks with 
even number of sequential layers in which WP propagates as
the second eigenstate. 

For large number of periods there are many elementary packets 
with the same values of 
$k_a, k_b, r, h$. We call this number the multiplicity, $M(k_a, k_b, r, h)$. 
Therefore the total amplitude of all WPs with a given 
$k_a, k_b$ equals
\begin{equation} \label{amp-kakb}
A(k_a, k_b) = \sum_{r,h} A(n,r,h) M (k_a, k_b,r,h).
\end{equation}


As for the one period case, here the shape of the total WP 
depends on mixing 
and multiplicities. 

For $k_a = k_a'$, $k_b = k_b'$ there is no shift but also 
the phase difference is zero. 
For $k_a - k_a' = 1$, $k_b - k_b' = 0$, the phase difference 
is $\phi^a$, while for $k_a - k_a' = 0$, $k_b - k_b' = 1$ the phase difference is $\phi^b$, {\it etc}. 
In general, one finds the components with phase differences varying from $0$ to $ n(\phi^a + \phi^b)$. 

\subsection{Resummation}
We can obtain the compact expression for the total WPs summing up effects after each crossing of the CW period representing them as the interference of the two (total)
WPs of $\nu_1^a$ and $\nu_2^a$.

We consider first the summation in the basis of eigenstates of the layer $a$:
$\nu^a = (\nu^a_1, ~\nu^a_2)$.
Let $\psi^{k - 1} = (\psi_1^{k - 1},~ \psi_2^{k - 1})^T$ be the vector describing
the state of system after crossing
$k - 1$ periods of the CW profile. $\psi_i^{k - 1}$ is result of resummation
of the elementary WP of $\nu_i^a$
in the layer $a$ in the beginning of $k$-period.
Then after crossing of the $k$th period, the state becomes
\begin{equation}
\psi^{k } = U_L^a \psi^{k - 1},
\label{eq:state-k}
\end{equation}
where the evolution matrix over one period 
\begin{equation}
U_L^a = U_\Delta D^b U_\Delta^{\dagger} D^a .
\label{eq:evmat}
\end{equation}
Here $U_\Delta$ is the matrix of the mixing change at crossing the border between $a$ and $b$:
\begin{equation}
U_\Delta = U^{a \dagger}U^b =
\left(\begin{array}{cc}
c_\Delta & s_\Delta \\
- s_\Delta & c_\Delta
\end{array}\right),
\label{eq:udelta}
\end{equation}
and $U^a$, $U^b$ are the flavor mixing matrices in the
$a$ and $b$ layers.
$D^a $ is the evolution matrix in the layer $a$ which can be written as
\begin{equation}
D^a =
\left(\begin{array}{cc}
S(t^a_1) e^{i2 \phi_1^a} & 0 \\
0 & S(t^a_2) e^{i2 \phi_2^a}.
\end{array}\right)
\label{eq:dop-a}
\end{equation}
Here $S(t_i^a)$ is the time-shift operator acting on the shape factor of the WP
in such a way that
$$
S(t_i^a) f_i (t) = f_i (t - t_i^a) , ~~~~ S(t_i^a) S(t_j^b) = S(t_j^b) S(t_i^a) = S(t_i^a + t_j^b) ~~~
(i, j = a, b).
$$
We can also require that $S(t)^{\dagger} = S(-t)$, so that $S$ is the unitary: $S(t)^{\dagger} S(t) = I$.

Similarly one can introduce the evolution matrix for the layer $b$. Inserting $U_\Delta$ and $D^m_i$ ($m = a, b$) from
(\ref{eq:udelta}) and (\ref{eq:dop-a}) into (\ref{eq:evmat}) we obtain
\begin{equation}
U_L^a =
\left(\begin{array}{cc}
c_\Delta^2 e^{-i\phi} + s_\Delta^2 S(t^b) e^{-i\phi'} & 
  s_\Delta c_\Delta [S(t^b + t^a) e^{i\phi} - S(t^a) e^{i\phi'} ] \\
s_\Delta c_\Delta [S(t^b) e^{- i\phi'} - e^{-i\phi} ] & 
  s_\Delta^2 S(t^a) e^{i\phi' } + c_\Delta^2 S(t^b + t^a) e^{ i\phi}
\end{array}\right),
\label{eq:evmat3}
\end{equation}
where we subtracted the commom phase factor $\exp \{i(\phi_2^b + \phi_2^a + \phi_1^b + \phi_1^a) \}$ and applied (\ref{phase-redefinition}) and (\ref{phase-redefinition-2}), 
as well as performed the time shift by 
$t_1^b + t_1^a$ and used (\ref{time-redefinition}). Using (\ref{eq:evmat3}), we can recover results in sect.~\ref{coherence and adiabaticity violation} for one period.


If all $S = 1$, we obtain the matrix which leads to the parametric oscillations as in \cite{Akhmedov:1988kd,Akhmedov:1998ui,Akhmedov:1999ty}.

The evolution matrix after $n$ periods equals
\begin{equation}
U_{nL}^a = (U_L^a)^n.
\label{eq:evmatnn}
\end{equation}
According to (\ref{eq:state-k}) the total eigenstate is
\begin{equation}
\psi^n (t) = (U^a_L)^n f(t) \psi^0,
\label{eq:eigen12}
\end{equation}
where $f(t)$ is the initial shape factor and
$\psi^0$ are the admixtures of eigenstates
$\nu_i^a$ in the initial moment of time.
The components of $\psi^n$ have the form
\begin{equation} \label{sum-kakb}
\psi^n= \sum_{k_a,k_b}^n A(k_a,k_b) f(t - k_a t^a - k_b t^b) e^{i (k_a \phi^a + k_b \phi^b) } \psi^0.
\end{equation}
Notice from (\ref{sum-kakb}) that oscillations in energy
are induced not only by interference between the total WPs of $\psi^n_1$ and $\psi^n_2$ but also by interference of components within amplitudes
$\psi_1^n$ and $\psi_2^n$ with amplitudes (\ref{amp-kakb}).

The amplitude of the $\nu_e \rightarrow \nu_e$ transition equals
\begin{equation}
A_{ee} (t) = \nu_e^T (U_L^a)^n f(t) \nu_e, ~~~~~~~~ \nu_e^T = (c_a,~ s_a).
\label{eq:ampl}
\end{equation}
This reproduces the amplitude for two layer case of sect.~\ref{parametric1}. Then the probability can be found as
\begin{equation}
P_{ee} = \int dt |\nu_e^T U_{nL}^a f(t) \nu_e|^2.
\label{eq:probab}
\end{equation}

Various results can be obtained using the general form of amplitude and probability. 
One can consider evolution in the flavor basis, which corresponds to the derivation of
the parametric oscillation probability in \cite{Akhmedov:1988kd,Akhmedov:1998ui,Akhmedov:1999ty}.
Now the evolution matrix equals
\begin{equation}
U_L^f = U^b D^b U_\Delta^{\dagger} D^a U^{a \dagger} = (U^b D^b U^{b \dagger})(U^a D^a U^{a \dagger}),
\label{eq:evmatf}
\end{equation}
where in the last step the eigenstates of the layer $b$ are projected onto the flavor basis.
The matrices in the $a$ eigenstate basis (\ref{eq:evmat}) and in the flavor basis (\ref{eq:evmatf})
are related as
\begin{equation}
U_L^a = U^{a \dagger} U_L^f U^{a}.
\label{eq:evmatfa}
\end{equation}
The matrices after $n$ layers:
\begin{equation}
(U_L^f)^n = U^{a} (U_L^a )^n U^{a \dagger}.
\label{eq:evmatfan}
\end{equation}


We can determine the effective Hamiltonian over one period using the evolution matrix over one period:
$U_L = I - i H L$. Then diagonalizing the Hamiltonian gives the effective mixing and the eigenvalues,
which, in turn, give the depth of the parametric oscillations and the phase.
Clearly, such a Hamiltonian depends on mixing $\Delta$ and the phases of individual layers
$\phi^a, \phi^b$.

\subsection{Effective group velocities and infinite coherence in CW
profile}

In general, the condition of infinite coherence can be formulated
as an equality of the effective
group velocities of the wave packets:
\begin{equation} \label{equal-eff}
v_1^{eff} = v_2^{eff}
\end{equation}
or $\Delta v^{eff} = 0$.
Determination $v_i^{eff}$ depends
on properties of the density profile and turns out to be non-trivial for
the CW profile.

Recall that in the case of constant density $v_i^{eff}$ are well defined, see (\ref{infcoh-conf}) and related discussion.
Since the shape factor does not change, the group velocity
is the velocity of any fixed point of shape factor, {\it e.g.}, maximum.
The difference of group velocities $\Delta v =
\Delta v (E, V, \theta, \Delta m^2)$
does not depend on the oscillation phase.
The velocities are constant and do not change in the course of
propagation.

If density changes along the neutrino trajectory,
$v_i^{eff}$ and $\Delta v^{eff}$ do depend on distance.
In the adiabatic periodically varying density
$v_i^{eff}$ can be introduced as the velocities averaged over period:
$$
v_i^{eff} = L^{-1} \int_0^L dx \frac{d H_i}{dE}\bigg|_{\bar{E}} .
$$
Then the condition of infinite coherence is
$$
\Delta v^{eff} = L^{-1}\int_0^L dx \frac{d \Delta H
}{dE}\bigg|_{\bar{E}} = 0.
$$
It corresponds to a situation when $\Delta v^{eff}(x) > 0$ in one part of the
period and $\Delta v^{eff}(x) < 0$ in another part (\ref{L_0}).

The situation is much more complicated if adiabaticity is broken as in the case of a single jump between two layers, discussed in sect.~\ref{coherence and adiabaticity violation}.
In the second layer, the WP has two components,
and it is non-trivial to identify the point of the WP to which the group velocity should
be
ascribed.
Furthermore, just after crossing the jump, certain parts of the WP will
have opposite phases and therefore interfere destructively,
so that the interference is the same as before crossing.
This also affects shape factors (\ref{total-WPs}) and (\ref{total-WPs-2}). In this case, the infinite
coherence has no meaning apart from specific cases which are reduced
to infinite coherence in a single layer (sect.~\ref{coherence and adiabaticity violation}).

Infinite coherence (infinite number of periods)
has meaning in the castle-wall profile for the
(long-range) parametric oscillations. The effective group velocities
should be defined for a single
period of CW profile similarly to the periodic profile
with adiabatic density change considered above. Furthermore,
one should take into
account (similarly to single jump case) that (\ref{sum-kakb})

\begin{itemize}

\item Shape factor of the total WPs changes at the crossings,
due to splitting and delays of the components;

\item Oscillation phase is different in different parts of the same
total WP because of different histories of the elementary components.

\end{itemize}

As a result, expressions for $v_i^{eff}$ and condition for
infinite coherence should depend not only on phases acquired in the
layers $a$ and $b$ but also on the mixing in both layers (\ref{eq:evmat3}).

Indeed we can obtain the condition in the $x$-representation rewriting condition
(\ref{divergency}) for $n_{coh} \rightarrow \infty$ in terms of
delays $t^a$ and $t^b$ we find:
\begin{multline}
\label{x-space}
 [s_a c_b +c_a s_b \cos 2\Delta] t_a +[c_a s_b+s_ac_b
\cos 2\Delta]t_b =\\
=s_a s_b \sin 2\Delta \frac{1}{E}
\left(\frac{V_{a} \sin 2 \theta_{a}}{\Delta H_{a}}-\frac{V_{b}
\sin 2 \theta_{b}}{\Delta H_{b}}\right),
\end{multline}
where we used
$$
\frac{d 2\theta_{m}}{dE}=\frac{V_{m} \sin 2 \theta_{m}}{E \Delta H_{m}},
$$
$m=a,b$. From (\ref{x-space}),
we see that infinite coherence for parametric oscillations
is unrelated to overlaps of the elementary WPs at the
end of each period; this is the major difference from the cases
where adiabaticity is conserved.

There is no simple interpretation of infinite coherence for parametric oscillations in terms of characteristics of elementary WP. In the
Table~\ref{table}, for three different $E_0$, we
show the delays of elementary WPs as well as phases and mixings in each layer of the CW profile with parameters (\ref{eq:paramet}). No simple correlation is observed.
\begin{table}
\begin{center} 
\begin{tabular}{ |p{1.8cm}||p{2cm}|p{2cm}|p{1.4cm}|p{1.4cm}| p{1.4cm}|p{1.4cm}| }
 \hline
 $E_0$ (MeV) & $t_a$ ($\text{MeV}^{-1}$) & $t_b$ ($\text{MeV}^{-1}$)&
$\phi^a$(rad) & $\phi^b$(rad) & $\theta_a$(rad) & $\theta_b$(rad) \\
 \hline
 $2.3$ & $-0.98$ & $-2.4$ & $1.6$ & $2.7$ &
$0.83$ & $0.15$ \\
 $3.3$ & $1.5$ & $-1.2$ & $2.1$ & $1.8$ &
$1.3$ & $0.156$ \\
 $7.7$ & $0.4$ & $-0.21$ & $3.87$ & $0.73$ &
$1.5$ & $0.2$ \\
 \hline
\end{tabular}
\caption{\label{table}Energies of enhanced coherence length, $E_0$, delays, phases and mixing angles in each of the layers $a$ and $b$ of a CW profile with parameters (\ref{eq:paramet}).}
\end{center}
\end{table}
\section{Applications to supernova neutrino evolution}
\label{applications}

The issues of coherence in matter are of great relevance for oscillations of supernova neutrinos (SN). 
The reasons are enormous distances from the production points to a detector at the Earth, 
complicated oscillation and conversion phenomena inside a star, non-trivial profiles of medium potentials. 
Here we briefly describe some effects, while a detailed study will be presented elsewhere \cite{yago:next}. 

SN neutrinos are produced at densities $\rho \sim 10^{12}$ $\text{g/cm}^3$ and have very short 
wave packets $\sigma_x \sim 10^{-11}$ cm \cite{Kersten:2013fba,Kersten:2015kio,Akhmedov:2017mcc}. Propagating to the surface of a star they can undergo 
the collective oscillations at $R < 100$ km \cite{Duan:2010bg,Mirizzi:2015eza,Chakraborty:2016lct,Tamborra:2020cul} and then 
the resonance flavor conversion at $R > 1000$ km \cite{Wolfenstein:1979ni,Mikheev:1986if,Dighe:1999bi,Lunardini:2003eh,Tomas:2004gr,Dasgupta:2005wn}. 
The collective effects are supposed to be active at certain 
phases of a neutrino burst \cite{Mirizzi:2015eza}. 

Let us first assume that collective oscillations are absent due to damping effects \cite{Raffelt:2010za} or 
due to lack of conditions for these oscillations at certain time intervals. 
In this case, the standard resonance conversion picture is realized: 
in the production region, the mixing is strongly suppressed so that the flavor states coincide with the eigenstates of propagation in matter. The coincidence depends on the type of mass hierarchy. 
Except for special situations (time periods and locations), the adiabaticity condition is well satisfied, 
and therefore the adiabatic transformations occur when neutrinos travel to the surface: $\nu_i^m \rightarrow \nu_i$
$(i = 1, 2, 3)$. The adiabaticity can be broken in fronts of shock waves which affects the described picture.

If the initial fluxes of $\nu_\mu$ and $\nu_\tau$ are equal, 
the results of conversion depend only on the $\nu_e - \nu_e$ survival probability, $p$, in the neutrino 
channel and on the $\bar{\nu}_e - \bar{\nu}_e$ probability, $\bar{p}$, in the antineutrino channel. 
For definiteness we will consider the $\Bar{\nu}_e$ channel. 
In this case, the flux of $\Bar{\nu}_e$ at the surface of the star, and consequently, the Earth equals 
\begin{equation}
 F_{\Bar{\nu}_e}=\Bar{p} F_{\Bar{\nu}_e}^0 + (1-\Bar{p}) F_{\Bar{\nu}_x}^0, 
\label{eq:eflux}
\end{equation}
where $F^0_{\Bar{\nu}_e}$ and $F^0_{\Bar{\nu}_x}$ are the initial fluxes of $\Bar{\nu}_e$ and $\Bar{\nu}_x$ 
(mixture of $\Bar{\nu}_\mu$ and $\Bar{\nu}_\tau$). 
The second term in (\ref{eq:eflux}) corresponds to transitions 
 $\Bar{\nu}_\mu, \Bar{\nu}_\tau \rightarrow \Bar{\nu}_e$. Derivation of the equation (\ref{eq:eflux}) assumes that 
initially produced $\nu_i^m$ (which coincide with the flavor states) are incoherent and evolve independently. 
Therefore signals produced by $\nu_i^m$ or states, to which $\nu_i^m$ evolve, 
sum up incoherently. 

For $R \ll 1000$ km, the matter density is much bigger than 
the density of the H-resonance associated with $\Delta m^2_{13}$. 
In this matter dominated range, according to Eq.~(\ref{matter-domination}) in 
sect.~\ref{coherence-lengths} the coherence length $L_{coh}^m$ is function of the vacuum parameters 
(see also fig.~\ref{fig-E0}): 
\begin{equation} \nonumber
L_{coh}^m = \frac{L_\nu}{\cos 2 \theta}\frac{E}{2\sigma_E} = 1100~ {\rm km} 
\left(\frac{E}{ 20 {\rm MeV }}\right) \left(\frac{E}{10 \sigma_E }\right).
\label{eq:cohinsn}
\end{equation}
Therefore the WP of eigenstates will separate before reaching the H-resonance even if they 
overlapped at the production. 
Notice that with an increase on $E$, the length $L_{coh}^m$ increases, but the resonance shifts to outer layers. Still, some partial coherence may exist.

For definiteness we assume the inverted mass hierarchy. Then according to the level crossing scheme 
at the production $\bar{\nu}_3^m \approx \bar{\nu}_e$, $\bar{\nu}_1^m \approx \bar{\nu}_x$ 
and $\bar{\nu}_2^m = \bar{\nu}_x'$. 
Furthermore, antineutrinos cross only H-resonance. 
The dynamics is related to $\bar{\nu}_3^m - \bar{\nu}_1^m$ subsystem, while 
$\bar{\nu}_2^m$ essentially decouples appearing as a spectator. Therefore the $\bar{\nu}_e$ survival probability can be written as 
\begin{equation} \nonumber
\label{survival}
\Bar{p}=|U_{e1}|^2 P_{31} + 
s_{13}^2 (1- P_{31}), 
\end{equation}
where $P_{31} \equiv P(\Bar{\nu}_{3}^m \rightarrow \Bar{\nu}_{1}^m)$ is the 
$\Bar{\nu}_{3}^m \rightarrow \Bar{\nu}_{1}^m$ transition probability.

Let us consider the effects of the coherence loss 
and coherence enhancement under different circumstances.

1. In the completely adiabatic case, $\bar{\nu}_3^m (\approx \bar{\nu}_e)$ evolves to $\bar{\nu}_3$, so that 
$P(\Bar{\nu}_{3}^m \rightarrow \Bar{\nu}_{1}^m) = 0$ and $\bar{p} = |U_{e3}|^2 = s_{13}^2$. 
In central parts of a star $\bar{\nu}_1^m$ has larger group velocity than $\bar{\nu}_3^m$. Below 
the $H$-resonance, 
inversely, $\bar{\nu}_3^m$ moves faster and $\bar{\nu}_3$ will arrive at the Earth first.

2. Adiabaticity is strongly broken in a shock wave front which can be considered as 
the instantaneous density jump (see sect.~\ref{coherence and adiabaticity violation}). 
We will assume that before the jump (layer $a$) and after the jump (layer $b$) neutrinos propagate adiabatically. 
Crossing the jump the eigenstates $\bar{\nu}_1^a$ and $\bar{\nu}_3^a$ split into eigenstates $\bar{\nu}_j^b$. 
To find $\bar{p}$ we consider evolution of $\bar{\nu}_3^a$ which becomes 
$c_\Delta \bar{\nu}_3^b - s_\Delta \bar{\nu}_1^b$ after crossing the jump. Then its components $\bar{\nu}_j^b$
loose coherence and evolve to $\bar{\nu}_j$ at the surface of a star. 
Consequently, we find 
\begin{equation}\nonumber
\Bar{p} = c_\Delta^2 s_{13}^2 + s_\Delta^2 |U_{e1}|^2. 
\label{eq:padviol}
\end{equation}
Strong adiabaticity violation effect on $\Bar{p}$ is realized if the change of mixing in matter, $\Delta$, is large. The latter occurs when the jump appears in the resonance layer, 
and the size of the jump is larger than the width of the resonance layer: 
$\Delta \rho > \rho_R \tan 2\theta_{13} = 0.3 \rho_R$. That is, even a small jump can produce a strong effect.

As we discussed before, the jump regenerates oscillations, which then disappear, 
and $\Bar{p}$ does not depend on the oscillation phase. Single shock front (jump) could 
lead to phase (coherence) effect, if $\bar{\nu}_1^a$ and $\bar{\nu}_3^a$ are produced in a coherent state 
({\it e.g.}, as a result of collective oscillations) and coherence is supported till the jump. 
In this case, the result of evolution depends on the phase $\phi^a$ acquired in the layer $a$, and this phase is observable. Notice that formula (\ref{eq:eflux}) is invalid in this case. \\

3. Let us consider the case of two density jumps that appear due to the presence 
of two shock wavefronts: the direct and reversed one. 
This situation is analogous to the one described 
in sect.~\ref{coherence and adiabaticity violation}. 
We denote by $a$, $b$ and $g$ the layers between the production point and the inner (reversed) shock, 
between two shocks and outside the outer shock correspondingly. 
The potential of the layer $b$, $V_b$ can be constant or change adiabatically, 
the potentials just before inner shock, $V_a$, and after second shock (jump), $V_g$, are not equal 
(in contrast to CW).

The picture of evolution of $\bar{\nu}_{3}^m$ state is the following:

\begin{itemize}

\item 
$\bar{\nu}_{3m}$ propagates adiabatically in the layer $a$ and splits at the border between $a$ and $b$: 
$\bar{\nu}_3^a = c_\Delta \bar{\nu}_3^b - s_\Delta \bar{\nu}_1^b$; 

\item
the latter state is coherent and oscillates in the layer $b$ acquiring the oscillation phase $\phi^b$: 
$c_\Delta e^{i\phi^b} \bar{\nu}_3^b - s_\Delta \bar{\nu}_1^b$; 

\item 
the eigenstates $\Bar{\nu}_3^b$, $\Bar{\nu}_1^b$ split in the second jump (the border between $b$ and $g$): 
$\bar{\nu}_3^b = c_{\Delta'} \bar{\nu}_3^g - s_{\Delta'} \bar{\nu}_1^g$, $\bar{\nu}_1^b = 
c_{\Delta'} \bar{\nu}_1^g + s_{\Delta'} \bar{\nu}_3^g$, 
where $\Delta' \equiv \theta_g - \theta_b$ is the change of the mixing angle in the second jump; 

\item 
in the layer $g$ the eigenstates $\nu_j^g$ evolve adiabatically to the mass eigenstates 
$\bar{\nu}_j^g \rightarrow \bar{\nu}_j$. Furthermore, $\bar{\nu}_1$ and $\bar{\nu}_3$ separate and decohere. 

\end{itemize}

According to this picture we obtain the following expression for the probability $\bar{p}$: 
\begin{equation}
\label{transition2}
\bar{p} = |U_{e1}|^2 \sin^2(\Delta + \Delta') + s_{13}^2 \cos^2(\Delta + \Delta')
-(|U_{e1}|^2 - s_{13}^2) \sin 2\Delta \sin 2\Delta' \sin^2\frac{\phi^b}{2}. 
\end{equation}
Notice that $\Delta + \Delta' = \theta_g -\theta_a$.

\begin{figure} 
\centering
\includegraphics[width=0.8\linewidth]{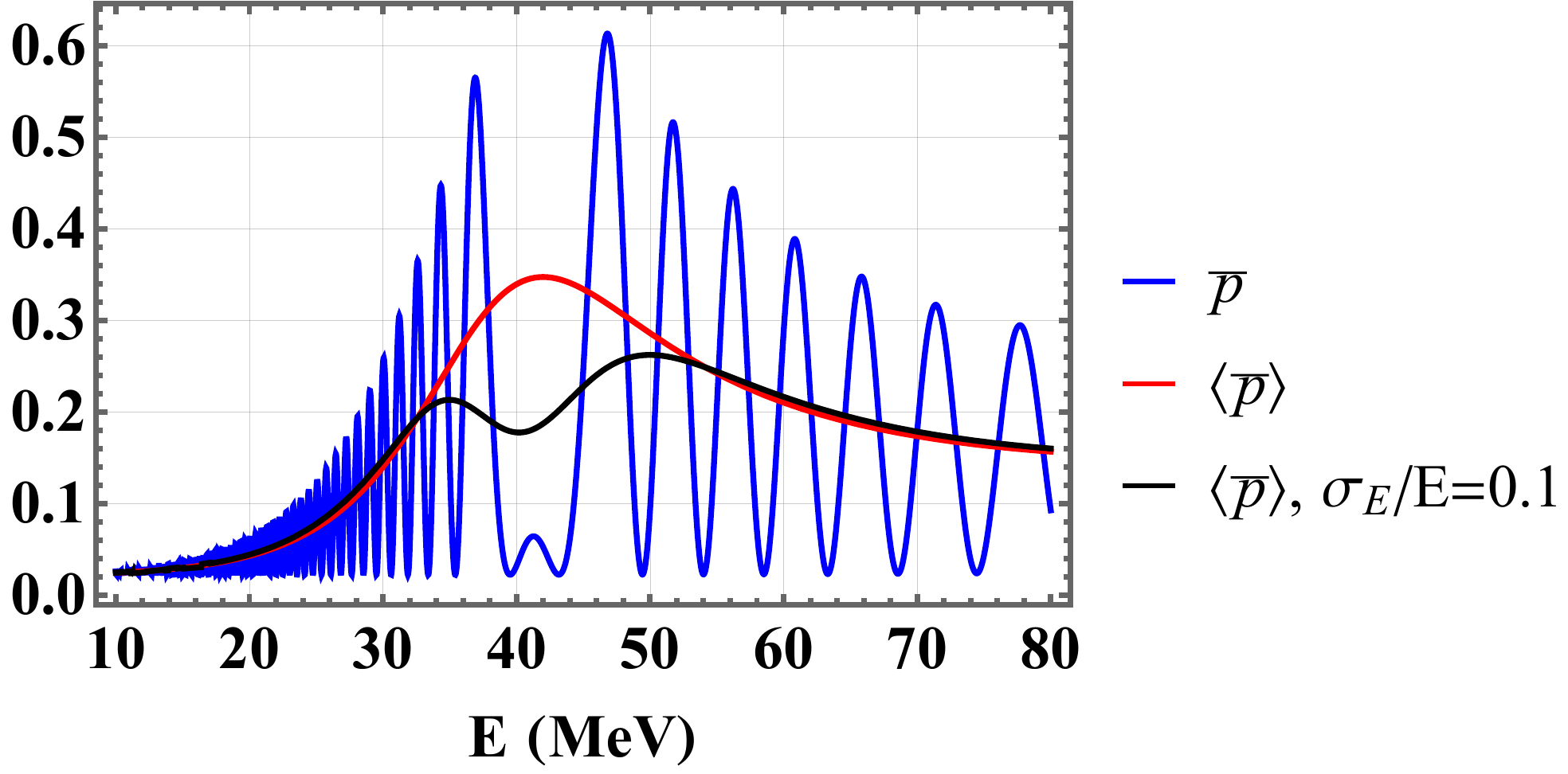}
\caption{Dependence of the $\bar{\nu}_e$ survival probability in supernova
on neutrino energy (blue line).
Inverted mass hierarchy and 1-3 vacuum mixing
were assumed. The probability
averaged over the energy interval
$\sigma_E = 0.1 E$ is shown by black line. The red line shows the probability without oscillatory
term.}
\label{dip}
\end{figure}

In fig.~\ref{dip} we plot $\Bar{p}$
computed with Eq. (\ref{transition2}) for the following values of 
parameters: $V_a=6.3 \times 10^{-12}$ $\text{eV}$, $V_b=3.2 \times 10^{-11}$ $\text{eV}$, $V_g=3.2 \times 10^{-12}$ $\text{eV}$ and $L_b=1060$ km. If we assume electron fraction $Y_e=0.5$, the corresponding densities are $\rho_a=2 \times 10^2$ $\text{g/cm}^3$, $\rho_b= 10^3$ $\text{g/cm}^3$, $\rho_g= 10^2$ $\text{g/cm}^3$.
Since the distance between the two shock fronts is 
much larger than the oscillation length in matter, $l_m < 100$ km, 
one can naively average the oscillatory terms in (\ref{transition2}), {\it i.e.} take 
$\sin^2 \phi^b/2 = 0.5$ (the 
red curve in fig.~\ref{dip}). However, according to our analysis in sect.~\ref{coherence-lengths} 
in layers of constant and slowly varying densities, the energy range of enhanced 
coherence exists around $E_0$, where oscillation effects might not be negligible. 
This is shown in fig.~\ref{dip} by the black curve for 
the energy resolution $\sigma_E/E=0.1$. Here $E_0=41$ MeV.
This oscillation effect will survive further propagation to the Earth, the effect encodes information about 
distance between shock wave fronts and density profile. Similar effects of enhanced coherence also survives for adiabatic propagation between the two shocks according to (\ref{eq:intcond3}).
As the distance between the two jumps increases, 
one needs better energy resolution, $\sigma_E/E < 0.1$, to observe the effect of enhanced coherence.

Let us consider the possible effect of the coherence loss on the collective
oscillations in central parts of a star.
These oscillations are induced by coherent flavor exchange in the $\nu - \nu$ scattering, and high neutrino densities at $r < 100$ km make them relevant.

The critical issue is whether fast loss of coherence due to
very short WPss destroys collective oscillations.
Indeed, the latter imply the $\nu - \nu$ interactions
after neutrino production, which can be considered as
$\nu$-detection ("observation") process.
Loss of coherence between production and interaction
as well as between different interactions can be important.
Furthermore, effectively the problem becomes
non-linear, and therefore, it is not clear if integration
over energies that form the WP can be interchanged with evolution.
Such an interchange is in the basis of equivalence 
of results in the $x$- and $E$-representations \cite{Stodolsky:1998tc}. There is no clear answer to 
this question \cite{Kersten:2013fba,Kersten:2015kio,Akhmedov:2016gzx,Akhmedov:2017mcc}.

Some new insight into the problem can be obtained using
description of the collective oscillations as
evolution of the individual neutrinos propagating
in effective external potentials formed by matter
and background neutrinos \cite{Hansen:2018apu}. In this case, the problem becomes
explicitly linear. Since the background
neutrinos oscillate, the effective potentials
have non-trivial oscillatory dependence in time (distance).
These variations of the potentials are non-adiabatic
and therefore strong flavor
transitions can be interpreted as parametric
oscillations and parametric enhancement effects. Taking into consideration this picture
one can apply the results of sect.~\ref{parametric-oscillations}.

The following comments are in order:

1. Propagation decoherence is related to the energy dependence
of the Hamiltonian (and consequently the phases). The collective (fast)
transformations are driven mainly by potentials that dominate
over the vacuum term $\Delta m^2/2E$ (the source of $E$ dependence).
The vacuum term (with mixing) triggers conversion at
the initial short phase. Therefore, apart from the initial phase,
 decoherence is simply absent.
If the initial phase is short enough
the decoherence is entirely irrelevant.

2. In the region of fast collective oscillations, $r_{coll} < 100$ km,
with strong matter dominance the coherence length
is given by vacuum parameters (\ref{eq:cohinsn}) 
and turns out to be $L_{coh} \sim 10^3$ km. So, $r_{coll} \ll L_{coh}$ and
decoherence can be neglected for $r < r_{coll}$.

3. For low energies ($E \sim 10$ MeV) and $\sigma_E \sim E$
the length $L_{coh} < 10^2$ km can be comparable to the
scale of collective oscillations.
However, this estimation of $L_{coh}$ is for constant density.
As we established in sect.~\ref{parametric-oscillations}, the parametric
oscillations themselves can substantially enhance coherence at least
in certain energy intervals around $E_0$. So, here we deal with
 coherence sustained by oscillatory dependence of
potentials.

Notice that consideration of coherence for collective oscillations
was problematic because of strong adiabaticity violation and
difficulty to define the eigenstates. Using analogy with the CW case, 
this can be done by integrating the Hamiltonian
oven one period, thus eliminating fast variations. For
the averaged Hamiltonian, one can introduce the eigenstates
and their effective group velocities.
Then the enhanced coherence corresponds to approximate equality
of these group velocities (\ref{x-space}).

\section{Conclusions} 
\label{conclusions}
The propagation decoherence occurs due to the difference
of the group velocities, which leads to the shift of
the WPs relative to each other and their eventual separation.
In matter, due to refraction, both the group
and phase velocities of neutrinos change
with respect to the vacuum velocities.
Thus, the refraction affects the propagation coherence (and
decoherence).

The consideration of WPs in the $x$-space
with average energy and momentum produces the same results
as consideration of plane waves in the $E$-space
and integration of probability (oscillation phase) over the energy
spectrum with width
given by the energy uncertainty.
In particular, the same value of the coherence length
follows from both approaches.

For different density profiles, we determined the coherence lengths
and their dependence on neutrino energy.
The salient feature in matter is the existence
of infinite coherence length,
$L_{coh} \rightarrow \infty$, at certain energies
$E_0$, and regions of enhanced coherence around $E_0$.
In the energy space, $E_0$ is given by zero derivatives
of the phase on energy. In the configuration space,
it follows from equality of the group velocities
of the eigenstates. The condition can be described as
the energy of minimal averaging of the interference term of the
probability.

The fundamental notion here is the effective group velocities
which depend on the density profiles.

1. For constant density the infinite coherence $E_0 = E_R/ \cos 2\theta$
coincides with the MSW resonance energy in oscillations
of mass eigenstates $\nu_1 \leftrightarrow \nu_2$, and for small vacuum mixing, it is close to the MSW flavor resonance.
The width of the region of enhanced coherence (weak averaging) is inversely proportional to the energy resolution, $\sigma_E$, and grows according to $\propto E_0^2$.
For very large densities (high energies), the coherence length
is determined by vacuum parameters and becomes close to the coherence
length in vacuum: $L_{coh}^m = L_{coh}/ \cos 2\theta$.

For massless neutrinos and oscillations driven by matter
the coherence is supported infinitely for all the energies.

2. In varying density profiles, the infinite coherence
is realized in particular situations.
For monotonously and adiabatically changing density
the coherence length can increase toward specific energy $E_0$,
which corresponds to certain $V_0$, such that for $V > V_0$ and
$V < V_0$ separation have opposite signs and compensate each other.

Infinite coherence can be realized
for periodic profile with adiabatic density change. It corresponds to
equality of the group velocities of eigenstates averaged over
the period. $E_0$ can be found from the latter condition.

3. In the presence of adiabaticity violation, transitions between
eigenstates
of propagation happen. This leads to modification of WP shape,
and consequently, to change of the effective group velocities.

We considered in details the case of maximal adiabaticity
breaking - density jumps in certain points of space (neutrino
trajectory).
In this case, two new effects are realized, which are related to
the split of the eigenstates at
borders between layers: (i) spread of total WP, (ii) change of the
oscillation phase
along these total WPs.

For the example of a single jump between two layers with different
constant densities,
there are two components in the first layer and four components in the
second.
Correspondingly there are various group velocities and oscillation
phases at a detector,
and it is not possible to introduce the coherence length in the usual
way. Still the medium
parameters can be selected so that some level of coherence can be
supported for long
distances because the split of the eigenstates regenerates interference
and oscillations.

4.
We checked that modifications of the total
WP in the $x$-space do not change its Fourier
transform to the $E$-space,
and consequently, the energy spectrum of neutrinos.
Therefore derivation of the infinite coherence
conditions for complicated density profiles can be more accessible in the
$E$-representation.

In certain situations, consideration of WP in the $x$-space
simplifies computations of probabilities, especially when
coherence is completely lost.
Moreover, it becomes crucial when time tagging
is introduced at the production or detection, breaking
the stationarity condition.

5. For periodic structures with density jumps, such as the castle-wall profile, the coherence length can be introduced for the parametric
oscillations
(and not for small scale coherence of elementary WP).
Then the coherence can be supported for many periods.
For parametric oscillations, there are several $E_0$, each associated with
a particular parametric resonance.

Interpretation of the condition for $E_0$ in $x$-space in terms of
elementary WPs propagating in the individual layers of the profile is very non-trivial.
We find that the effective group velocities for one period (and
therefore the
infinite coherence condition) are rather complicated functions of mixing angles, phases and delays acquired by the elementary WPs in two parts of the period.

6. Using elementary WPs, we reconstructed total WPs at a detector and
studied their spread and shape change
in the course of propagation depending on parameters of the CW profile. The reconstruction of the total WP
from the elementary WP using operators of time shift is presented. Resummation of the elementary WP at a detector can be
performed using the evolution matrix for one period
expressed in terms of operators of time shifts.

7. We outlined applications of the obtained results to supernova neutrinos.
In particular, we showed that coherence can be supported between
two shock wavefronts, leading to observable
oscillation effects at the Earth.

For collective oscillations 
due to large matter potentials
the coherence is determined by vacuum parameters and
coherence length is much larger than the scale of
fast oscillations. Using interpretation
of collective oscillations as the parametric effects
we find that coherence can be further enhanced in certain
energy ranges.

\section{Acknowledgements}
YPPS acknowledges support from FAPESP funding Grants No. 2014/19164-453 6, No. 2017/05515-0 and No. 2019/22961-9.

\bibliographystyle{unsrt}
\bibliography{refs.bib}
\end{document}